\documentclass[article,shortnames,nojss]{jss}

\usepackage{amsfonts}
\newcommand{\Real}{\mathbb{R}}

\author{Adelchi Azzalini\\ Universit\`a di Padova  
     \And 
       Giovanna Menardi \\ Universit\`a di Padova  }
\title{Clustering Via Nonparametric Density Estimation: 
     the \proglang{R} Package \pkg{pdfCluster} }

\Plainauthor{Adelchi Azzalini, Giovanna Menardi}  
\Plaintitle{Clustering Via Nonparametric Density Estimation: %
   the R Package pdfCluster}
\Shorttitle{Clustering Via Nonparametric Density Estimation}  

\Abstract{
The \proglang{R} package \pkg{pdfCluster} performs cluster analysis based on 
a nonparametric estimate of the density of the observed variables. After 
summarizing the main aspects of the methodology, we describe the features and
the usage of the package, and finally illustrate its working with the aid of two datasets.
}
\Keywords{cluster analysis, nonparametric density estimation, kernel
  methods, \proglang{R}, unsupervised classification,  Delaunay triangulation}
\Plainkeywords{ cluster analysis, nonparametric density estimation, kernel
  methods, R, unsupervised classification,  Delaunay triangulation} 

\Address{
  Adelchi Azzalini\\
  Dipartimento di Scienze Statistiche\\
  Universit\`a di Padova, Italia\\
  E-mail: \email{azzalini@stat.unipd.it}\\
  URL: \url{http://azzalini.stat.unipd.it} 
 \par\vspace{1ex}
  Giovanna Menardi\\
  Dipartimento di Scienze Statistiche\\
  Universit\`a di Padova, Italia\\
  E-mail: \email{menardi@stat.unipd.it}
}
\begin{document}
 

\section{Clustering and density estimation}

The traditional approach to the clustering problem, also called `unsupervised
classification' in the machine learning literature, hinges on some notion of
distance or dissimilarity between objects. Once one such notion has been
adopted among the many existing alternatives, the clustering process aims at
grouping together objects having small dissimilarity, placing instead those
with large dissimilarity in different groups.  The classical reference for
this approach is \citet{H75}; the current standard account is
\citet{Kauf:Rous:1990}.

In more recent years, substantial work has been directed to the idea that the
objects to be clustered represent a sample from some $d$-dimensional random
variable $X$ and the clustering process can consequently be carried out on the
basis of the distribution of $X$, to be estimated from the data themselves.
It is usually assumed that $X$ is of continuous type; denote its density
function by $f(x)$, for $x\in\Real^d$. We shall discuss later how the
assumption that $X$ is continuous can be circumvented.

The above broad scheme can be developed in at least two distinct directions.
The first one regards $X$ as a mixture of, say, $J$ subpopulations, 
so that its density function takes the form
\begin{equation} \label{f:mixture}
           f(x) = \sum_{j=1}^J \pi_j\,f_j(x) 
\end{equation}
where $f_j$ denotes the density function of the $j$-th subpopulation and
$\pi_j$ represents its relative weight; here $\pi_j>0$ and $\sum_j \pi_j=1$.
In this logic, a cluster is associated to each component $f_j$ and any given
observation $x'$ is allocated to the cluster with maximal density $f_j(x')$
among the $J$ components. 

To render the estimation problem identifiable, some restriction must be placed on
the $f_j$'s. This is tipically achieved by assuming some parametric form for
the component densities, whence the term `model-based clustering' for this
formulation, which largely overlaps with the notion of finite mixture
modelling of a distribution. Quite naturally, the most common assumption for
the $f_j$'s is of Gaussian type, but other families have also been considered.
The estimation problem now involves estimation of the $\pi_j$'s and the set of
parameters which identify each of $f_1,\dots,f_J$ within the adopted
parametric class.  An extended treatment of finite mixture models is given by
\citet{Mclachlan_Peel00}.
 
There exist some variants of the above mixture approach, but we do not dwell
on them, since our main focus of interest is in the alternative direction which
places the notion of density function in a nonparametric context. The chief
motivation for this choice is to free the individual clusters from a given
density shape, that is the parametric class adopted for the components $f_j$
in Equation~\ref{f:mixture}.  If the cluster shapes do not match the ones of the
$f_j$'s, the mixture approach may face difficulties. This problem can be
alleviated by adopting a parametric family of distributions more flexible than
the Gaussian one. For instance, \citet{LinTI:LeeJC:HsiehWJ:2007} adopt the
skew-$t$ distribution for the $f_j$ components; this family provides better
adaptability to the data behaviour, and correspondingly can lead to a reduced 
number $J$ of components,  compared to the Gaussian assumption.
Although variants of this type certainly increase the flexibility of the
approach, there is still motivation for considering a nonparametric formulation,
completely free from assumptions on the cluster shapes.

The alternative approach to the use of density function in clustering places
then $f(x)$ in a nonparametric context. Since this direction has been examined
relatively more recently than the parametric one, it is undoubtedly less
developed, but growing.  It is not our purpose here to provide a systematic
review of this literature, especially in the present setting, considering
that very few of the methodologies proposed so far have lead to the
construction of a broadly-usable software.
We restrict ourselves to mention the works of \citet{Stuetzle03} and 
 \citet{Stuetzle_Nugent10}. In the supplementary material provided by this latter reference,
 the \proglang{R} package \pkg{gslclust} is also available.
Among the few ready-to-use techniques, a quite popular one is `dbscan' by
\citet{DBSCAN96}, originated in the machine learning literature and available through 
the \proglang{R} package \pkg{fpc} \citep{Fpc}; the notion of
data density which they adopt is somewhat different from the one of probability theory
considered here. For more
information on the existing contributions in this stream of literature, we
refer the reader to the discussion included in the papers to be summarized in
the following section.

It is appropriate to clarify that the above two approaches involve somewhat
different notions of cluster. In the nonparametric context, clusters are
associated to regions with high density, while in the parametric setting
(\ref{f:mixture}) they are associated to the components $f_j$. The two concepts
are different, even if they often lead effectively to the same outcome. A
typical case where they diverge is provided by the mixture of two bivariate
normal populations, both with markedly non-spherical distribution, such that
where their tails overlap an additional mode, besides the centres of two
normal components, is generated by the superposition of the densities; see for
instance Figure~1 of \citet{Ray_Lindsay2005} for a graphical illustration of
this situation. In this case, the mixture model (\ref{f:mixture}) declares
that two clusters exist, while from the nonparametric viewpoint, where
subpopulations do not feature, the three modes translates into three clusters.

The present paper focuses on the  clustering methodology constructed via
nonparametric density estimation developed by \citet{AT} and by 
\citet{Menardi_Azzalini2012}. Of this formulation, we first recall
the main components of the methodology and then describe its \proglang{R}
implementation \citep{R} in the package \pkg{pdfCluster} \citep{PdfClusterpkg}, illustrated with some
numerical examples.

\section{Clustering via nonparametric density estimation}

\subsection{Basic notions}

The idea of associating clusters to modes or to regions of high density goes
back a long time. \citet{Wishart69} stated that clustering methods should
be able to identify ``distinct data modes, independently of their shapes
and variance''. \citet[p.\,205]{H75} stated that
  ``clusters may be thought of as regions of high density separated
  from other such regions by regions of low density'',
and the subsequent passage expanded somewhat this point by considering
`density-contour' clusters formed by regions with density above a given
threshold $c$, and showing that these regions form a tree as $c$ varies.
However, this direction was left unexplored and the rest of Hartigan's book,
as well as the subsequent mainstream literature, developed cluster analysis methodology
in another direction, that is building on the notion of dissimilarity.

Among the few subsequent efforts to build clustering methods based on the idea
of density function in a nonparametric context, we concentrate on the
construction of \citet{AT} and its development of
\citet{Menardi_Azzalini2012}, which we summarize up to some minor differences.

For a $d$-dimensional density function $f(\cdot)$, which we assume to satisfy
adequate regularity conditions, define
\begin{eqnarray}
   R(c) &=& \{x: x\in\Real^d, f(x) \geq c \}, \qquad (0 \le c\le \max f),
              \label{R(c)} \\
   p_c &=& \int_{R(c)} f(x) \,\mathrm{d}x  \nonumber
\end{eqnarray}
which represent the region with density values above a level $c$ and its
probability, respectively.  For any given $c$, $R(c)$ may be a connected set
or not; in the latter case, we have detected the presence of two or more
high-density regions. 

\begin{figure}
  \centerline{\includegraphics[width=0.5\hsize]{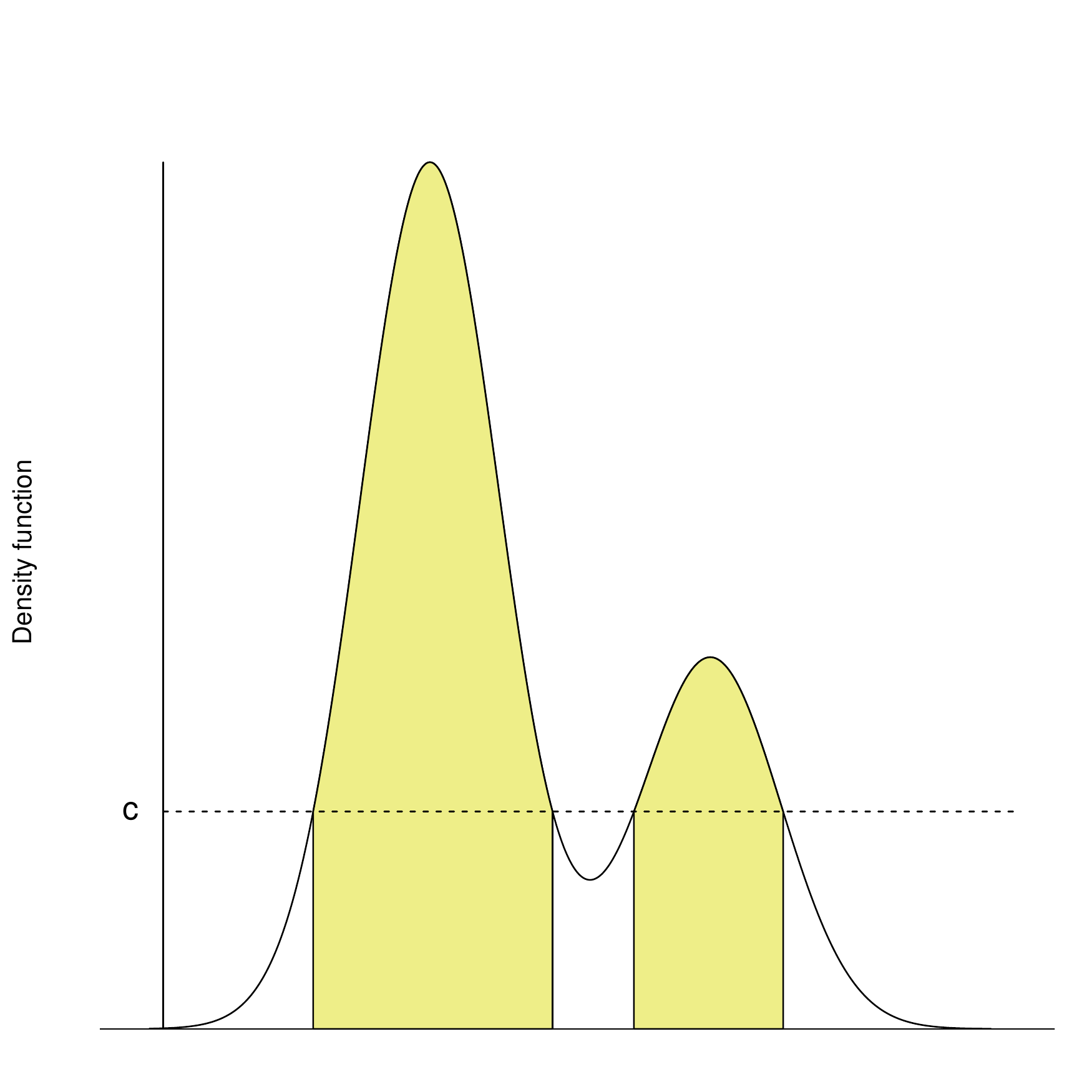}\hfill
    \includegraphics[width=0.45\hsize]{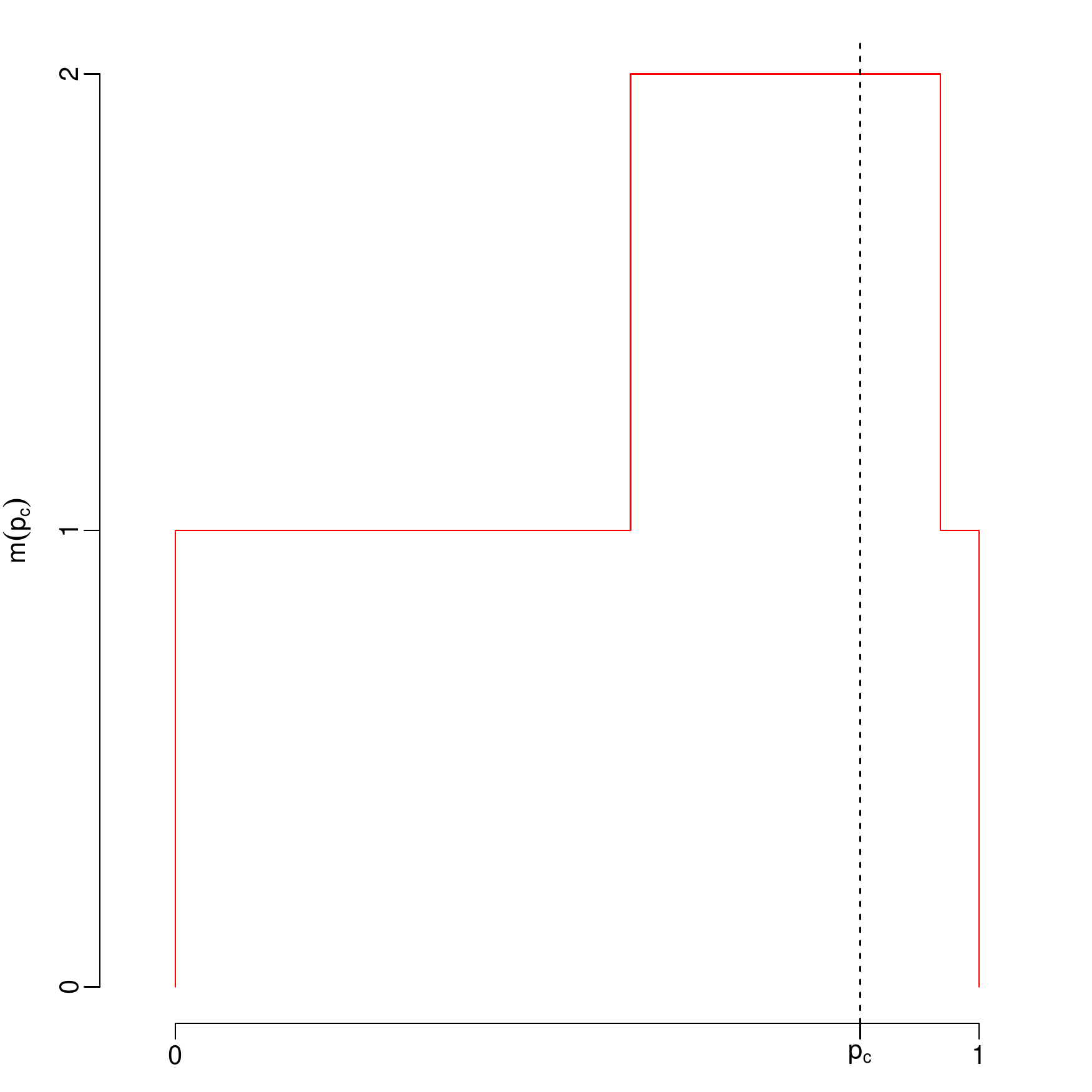}}
 \caption{Density function and set $R(c)$ for a given $c$ (left panel) and
    corresponding mode function (right panel).}
  \label{fig:pdf}
\end{figure}

These notions are illustrated for the case $d=1$ by the left panel of
Figure~\ref{fig:pdf}, where the two intervals at the basis of the shaded area
jointly represent $R(c)$ and the area itself represents $p_c$, for a specific
choice of $c$. As $c$ varies, the number of connected regions varies. Since
$c$ and $p_c$ are monotonically related, we can regard the number of
connected regions as a step function $m(p)$ of $p$, for $0<p<1$; for
convenience, we set $m(0)=m(1)=0$.  The right panel of Figure~\ref{fig:pdf}
displays the function $m(p)$ corresponding to the density of the left panel.

We shall refer to $m(p)$ as the `mode function'  because it enjoyes some 
useful properties related to the modes of $f(x)$. The most relevant facts are: 
(a)~the total number of increments of $m(p)$, counted with their multiplicity, 
is equal to the number of modes, $M$; 
(b)~a similar statement holds for the number of decrements;
(c)~the increment of $m(p)$ at a given point $p$ equals the number
of modes whose ordinate is $c_p$. Inspection of the mode function allows
to see, moving along the $p$ axis, when a new mode is given `birth', 
or when two or more disconnected high-density sets merge into one.

Moreover, as established by \citet[Section 11.13]{H75}, the set of regions
$R(c)$ exhibits a hierarchical structure. This tree structure
will be illustrated later in the examples to follow.

When a set of observations $S=\{x_1,\dots, x_n\}$ randomly sampled from $X$ is
available, we can compute a nonparametric estimate $\hat{f}(x)$ of the
density. The specific choice of the type of estimate is not crucial at this
point, provided it satisfies commonly required properties of nonparametric
estimates. The sample version $\hat{R}(c)$ of $R(c)$ is then obtained
replacing $f(x)$ by $\hat{f}(x)$ in Equation~\ref{R(c)}, and a corresponding sample
version of the mode function is introduced. Under mild regularity conditions,
one can prove `strong set consistency' of $\hat{R}(c)$ to $R(c)$, as
$n\to\infty$.

Since we are primarity interested in allocating observations to clusters, far 
more often than allocating all points of $\Real^d$, this can be achieved
considering
\begin{eqnarray}
        S(c) &=& \{x_i: x_i\in S, \hat{f}(x_i)\geq c \}, \qquad 
                 (0 \le c\le \max \hat{f}),    \label{S(c)}\\
   \hat{p}_c &=& |S(c)|/n \nonumber
\end{eqnarray}
where $|\cdot|$ denotes cardinality of a set. Again, under mild conditions,
one can show that $\hat{p}_c\to p_c$ as $n\to\infty$.

The above construction is conceptually simple and clear, but its actual
implementation is problematic. While for $d=1$ identification of $R(c)$ and of
$S(c)$ is elementary, as perceivable from Figure~\ref{fig:pdf}, the problem
complicates substantially for $d>1$, which of course is the really interesting
case. More specifically, it is immediate to state whether any given point $x$
belongs to any given set $R(c)$, but it is harder to say how many connected sets
comprise $R(c)$, and which one they are; a similar problem exists for $S(c)$.
The next two sections describe two routes to tackle this question.

\subsection{Spatial tessellation}\label{sec:delaunay}

To establish whether a set $S(c)$ is formed by points belonging to one or more
connected regions which comprise $\hat{R}(c)$, \citet{AT} make use of some
concepts from computational geometry.  The first of these is the Voronoi
tessellation which partitions the Euclidean space into $n$ polyhedra, possibly
unbounded, each formed by those points of $\Real^d$ which are closer to one given
point in $S$ than to the others.  Conceptually, from here one derives the Delaunay 
triangulation which is formed by joining points of $S$ which share a facet in the Voronoi 
tessellation.  From a computational viewpoint, the Delaunay triangulation can be obtained 
directly, without forming the Voronoi tessellation first, and it is the only one required 
for the subsequent steps. The elements of the Delaunay triangulation are simplices in 
$\Real^d$; since for $d=2$ they reduce to triangles, this explain the term.

\begin{figure}
  \centerline{
    \includegraphics[width=0.43\hsize]{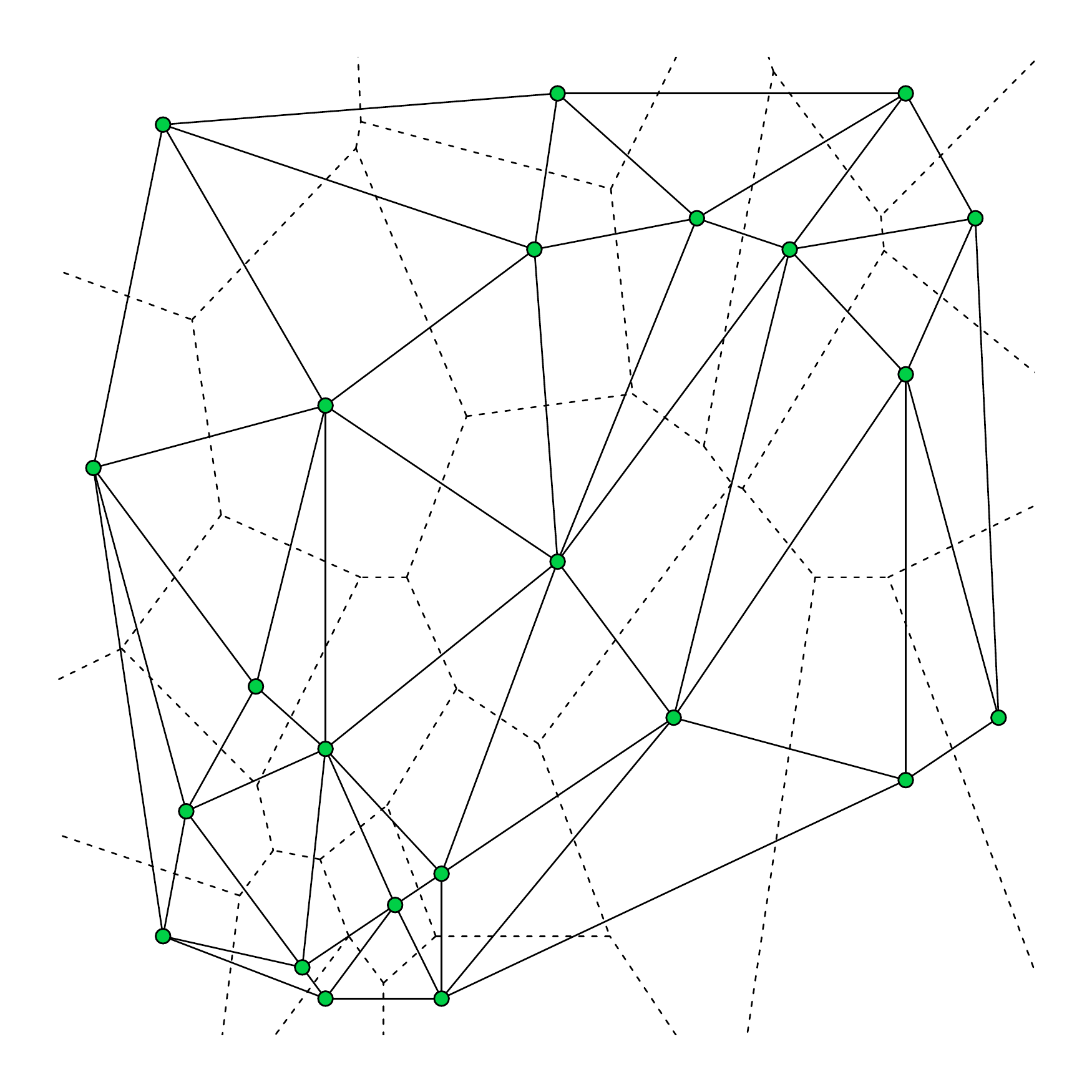}\hfill
    \includegraphics[width=0.43\hsize]{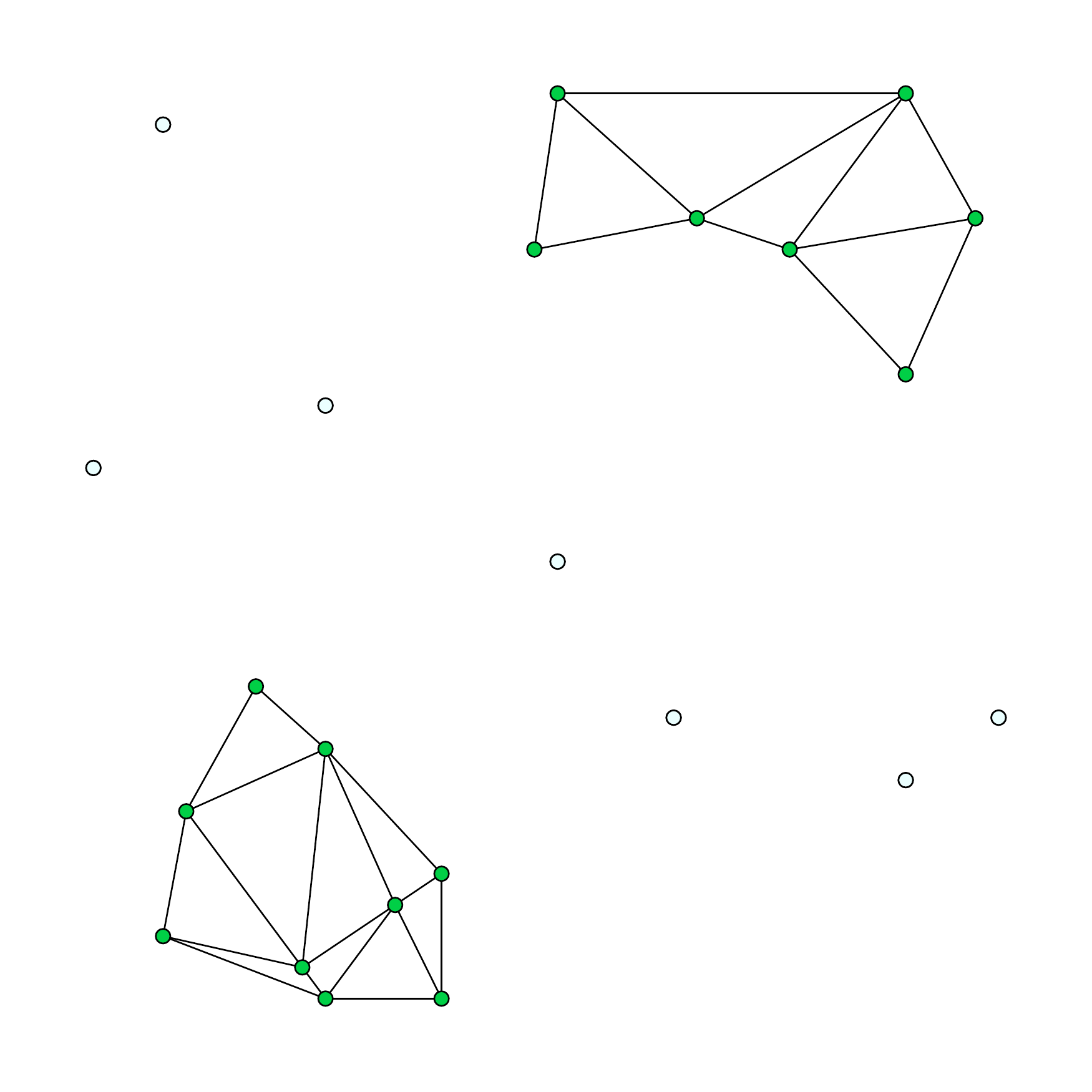}}
  \caption{The left plot displays an example of Voronoi
    tessellation (dashed lines) for a set of points when $d=2$, and
    superimposed Delaunay triangulation (continuous lines).  The right
    plot removes edges of some points from the original Delaunay
    triangulation, keeping points with $\hat{f}>c$ for some threshold
    $c$.}
  \label{fig:DT}
\end{figure}

These notions are illustrated in left panel of Figure~\ref{fig:DT} which shows
the Voronoi tessellation and the Delaunay triangulation, 
for a set of points in $\Real^2$.

The procedure of \citet{AT} consists of two main stages. The first one
comprises itself a few steps, as follows.  First, we construct the Delaunay
triangulation of the sample $S$, and compute the nonparametric estimate
$\hat{f}(x_i)$ for each $x_i\in S$. Then, for any given value $p_c\in(0,1)$,
we eliminate all points $x_i$ such that $\hat{f}(x_i)<c$ and determine the
connected sets of the remaining points.  This step is illustrated graphically
in the right panel of Figure~\ref{fig:DT}, where two connected sets are
visible after removing the points of low density and the associated arcs from
the triangulation on the left panel.  In principle, this operation is repeated
for all possible values $p_c\in(0,1)$, in practice for a grid of such points.
At the end of this process, we can construct a tree of these connected sets,
provided we have kept track of the group membership of the sample components
as $p_c$ ranges from $0$ to $1$. In the same process, we have also singled out
$M$, say, groups of points which form the connected sets so identified; we
call them `cluster cores'.  It is a quite distinctive feature of this method
to pinpoint a number $M$ of groups, while most methods require that $M$ is
specified on input or it is left undetermined, like in hierarchical
distance-based methods.

\begin{figure}
  \centerline{
    \includegraphics[width=0.46\hsize]{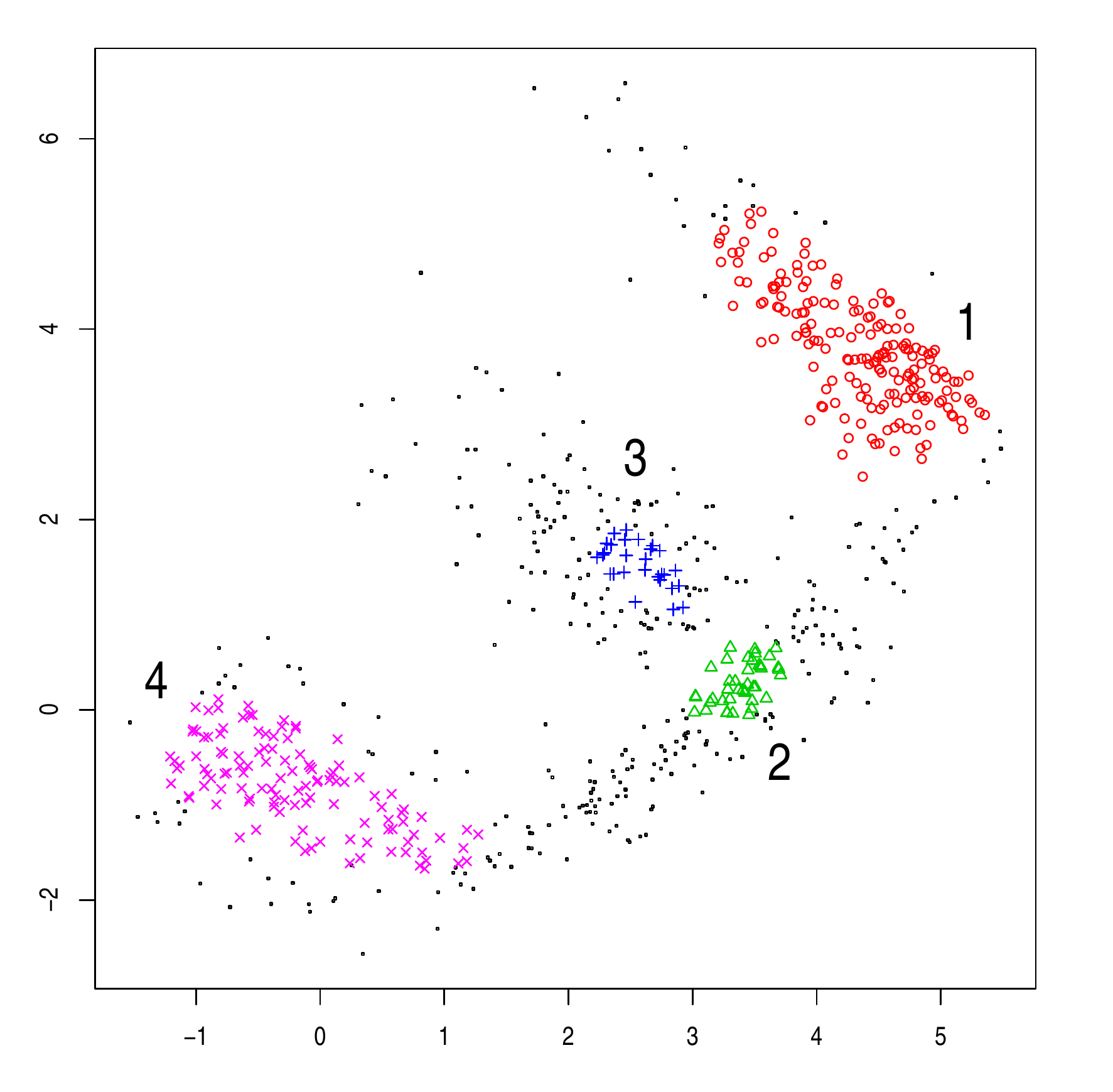}
    \hfill
    \includegraphics[width=0.46\hsize]{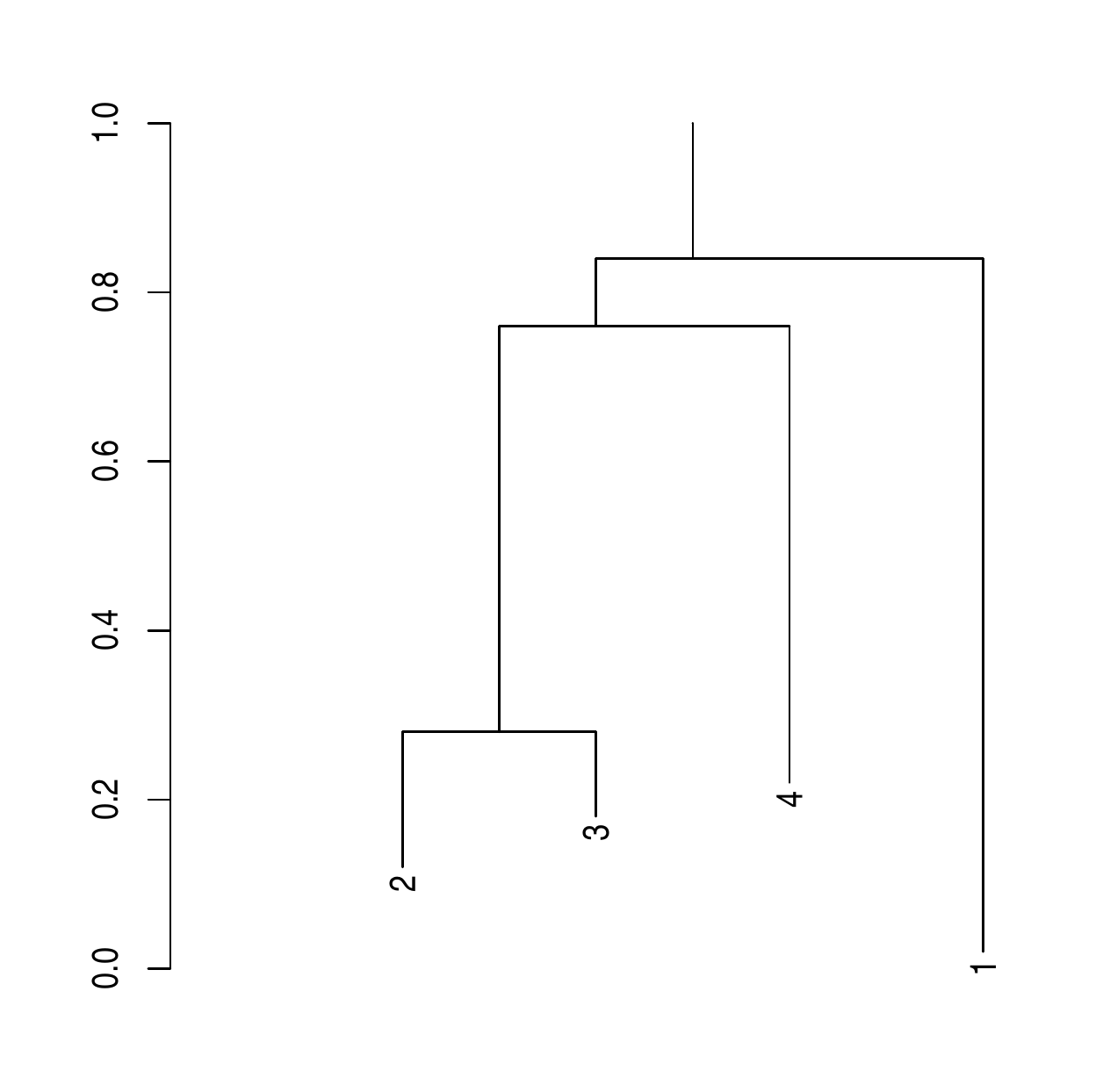}
  }
  \caption{Cluster cores and cluster tree for a set of simulated data with
    $d=2$. In the left panel the points belonging to the custer chores are
    marked as follows: $1=\mathrm{o}$, $2=\triangle$, $3=+$, $4=\times$; the
    unlabelled points are marked by dots. The right panel shows the
    corresponding cluster tree.}
  \label{fig:simul3}
\end{figure}

The outcome of the first stage is illustrated in Figure~\ref{fig:simul3} for a
set of simulated data with $d=2$. The left panel displays the observations
points, marked with different symbols for the four cluster cores; the
unlabelled points are denoted by a simple dot. The right panel shows the
cluster tree of the four groups. Notice that the tree is upside-down with
respect to the density: the root of the tree corresponds to zero density and
the top mode, marked by 1, is near the bottom. This confirms that it would
make more sense if mathematical trees had they roots at the bottom, like real
trees.

In the second stage of the procedure, the points still unlabelled must be
allocated among the $M$ cluster cores. This operation is a sort of
classification problem, although of a peculiar type, since the unlabelled
points are not randomly sampled from $X$, but inevitably in the outskirts of
the cluster cores.  Among the many alternative options, the proposed criterion
is based on density estimation and density ratios.  For an unlabelled point
$x_0$, compute an estimate $\hat{f}_m(x_0)$ of its density with respect to
each of the cluster cores, that is for $m=1,\dots,M$, and assign $x_0$ to the
group with highest log-ratio
\begin{equation}  \label{pdf-ratio}
     r_m(x_0) = \log \frac{\hat{f}_m(x_0)}{\max_{k\not=m}\hat{f}_k(x_0)}\,.
\end{equation}
It is also proposed not to compute the estimates of the cluster cores densities
$\hat{f}_m(\cdot)$ once for all, but to update them in a block-sequential
matter, after a certain fraction of points has been allocated, and so on
repeatedly for a given number of such blocks.


A variant of this allocation rule, not examined by \citet{AT}, but which
we have found preferable on the whole, weights the the log-ratios  in Equation~\ref{pdf-ratio} inversely with their variability while defining the order of allocation of the low density data. In practice, computation of the standard error of (\ref{pdf-ratio}) is quite complicated, even employing asymptotic expressions of variances. 
We take a rough approximation of those quantities, instead, by first identifying the index 
$m'$ such that $\hat{f}_{m'}(x_0)=\max_{k\not=m}\hat{f}_k(x_0)$ and then considering the 
$m'$ index as given. Next, we apply standard approximations for transformation of variables; 
see \citet[][page 29]{Bowman_Azzalini97}, for the specific case of
transforming $\hat{f}(\cdot)$.

As a diagnostic device to evaluate the quality of the clusters so obtained,
 the density-based silhouette (\emph{dbs})
proposed by \citet{Menardi:2011} fits naturally in this framework.
This tool is the analogue of the classical silhouette information  
\citep{Rousseeuw_87}, when the distances among points are replaced by probability log-ratios. 
Specifically, on defining for observation $x_i$,
\[ 
  \hat\tau_m(x_i)= 
     \frac{\pi_{m}\hat{f}_m(x_i)}{\sum_{k=1}^M \pi_{k}\hat{f}_k(x_i)},
   \qquad m=1,\ldots,M,
\] 
where $\pi_m$ is a prior probability for the $m$-th group,
the \emph{dbs} for $x_i$ is
\[
\mbox{\emph{dbs}}_i =\frac{\log
  \left(\frac{\hat{\tau}_{m_{0}}(x_i)}{\hat{\tau}_{m_{1}}(x_i)}\right)}%
  {{\max}_{x_i  }\left|  \log\left(\frac{\hat{\tau}_{m_{0}}(x_i)}%
    {\hat{\tau}_{m_{1}}(x_i)}\right)\right|},
\]
where $m_0$ is the group to which $x_i$ has been allocated and $m_1$ is
the alternative group for which $\tau_m(x_i)$ is maximal. 
The  interpretation of the \emph{dbs} diagnostics is the same as for the 
classical `silhouette'. 

We close this section with some remarks on computational aspects. From the 
point of view of memory usage, the requirement of this procedure grows linearly 
with $n$, while for methods based on dissimilarities it grows quadratically.
The more critical aspect here is construction of the Delaunay triangulation.
This can be produced by the Quickhull algorithm by \citet{Barber_Etal96}, 
whose implementation is publicly available at 
\url{http://www.qhull.org}.  This algorithm works efficiently for increasing 
$n$ when $d$ is small to moderate, but computing time grows exponentially
when $d$ increases. 

\subsection{Pairwise connections}\label{sec:pairs}

The final remarks of the previous section motivate the development of a
variant procedure to build a connection network of the elements of $S$
by using a different criterion instead of the Delaunay triangulation,
leaving unchanged the rest of the above-described process. 

The proposal of \citet{Menardi_Azzalini2012} starts by reconsidering the
notion of connected sets for $d=1$  and then extending this view to the 
case $d>1$.  The basic idea is to examine the behaviour of $\hat{f}(x)$
when we move along a segment $[x_1, x_2]$, since it depends on whether the sample
values $x_1$ and $x_2$ belong to the same connected set of $\hat{R}(c)$ or not.
To visualize the process, it is convenient to refer to the left panel of 
Figure~\ref{fig:pdf}, regarding the density there as the estimate $\hat{f}$, 
and consider the set $\hat{R}(c)$ formed by the union of the two intervals 
of the shaded area. 
If $x_1$ and $x_2$ belong to the same interval, then  
the corresponding portion of density along the segment $(x_1, x_2)$ has no 
local minimum. 
On the contrary, if $x_1$ and $x_2$ belong to different subsets of 
$\hat{R}(c)$, then at some point along $[x_1, x_2]$ the density 
exhibits a local minimum, which we shall refer to as `presence of a valley'.

When $d>1$, the same idea can be carried over by examining the behaviour of
the section of $\hat{f}(x)$ along the segment joining $x_1$ and $x_2$ 
where $x_1, x_2\in\Real^d$  and applying the same principle as above. 
A graph is  then created whose vertices are
the sample points and an arc is set between any pair of points such that
there is no valley in the section of $\hat{f}(x)$ between them.

In practical terms, the claim of presence of a valley must allow some tolerance
for the inevitable variability of $\hat{f}$. Given the above
premises, we must take into account the amplitude of the valley detected along
the stated section of $\hat{f}$, and declare that $x_1$ and $x_2$ are connected
points if this amplitude is below a certain threshold. Clearly, if no valley
exists, the connection of $x_1$ and $x_2$ is unquestioned.

\begin{figure}
  \centerline{
    \includegraphics[width=0.5\hsize]{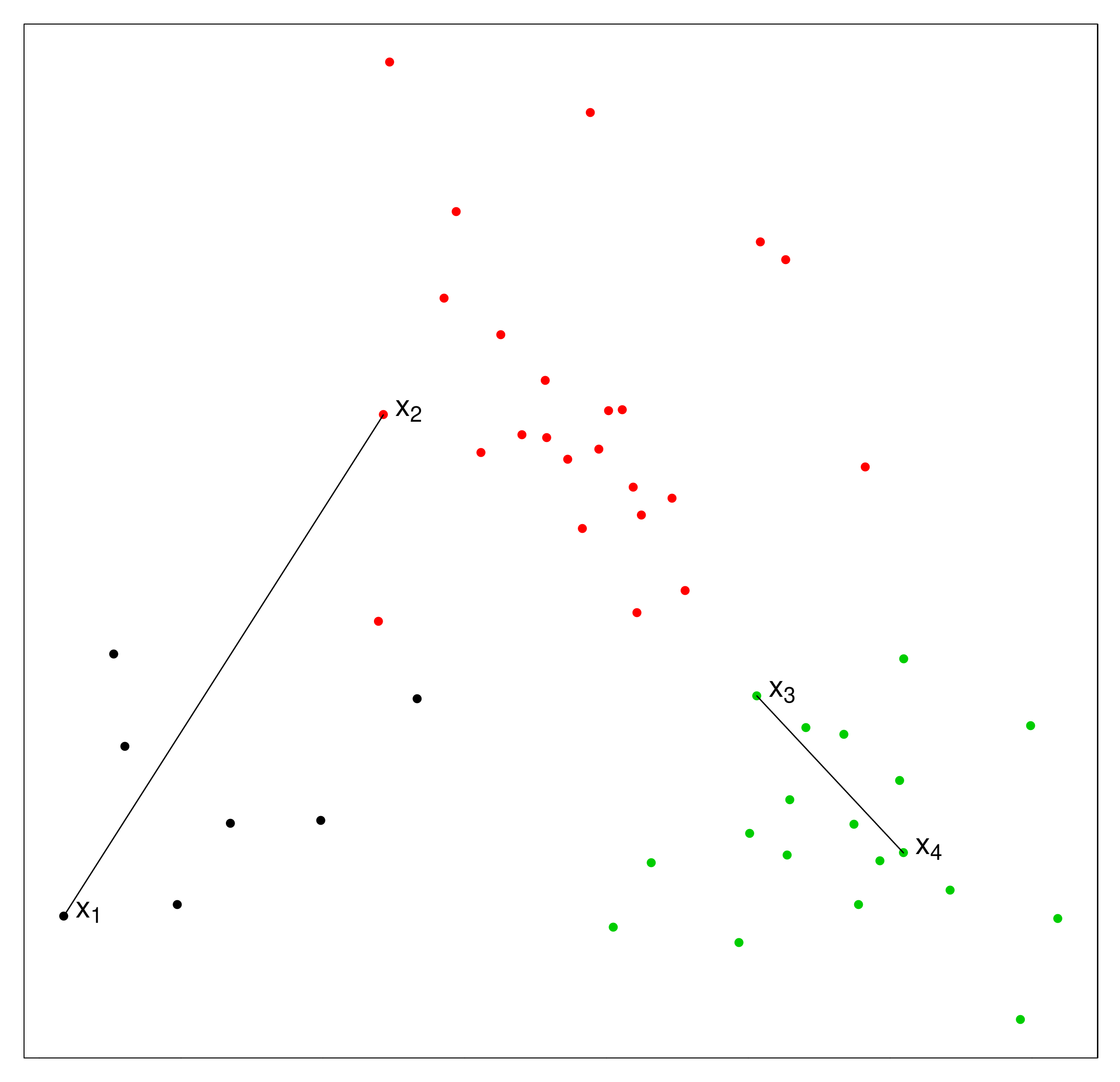}
    \hfill
    \includegraphics[width=0.5\hsize]{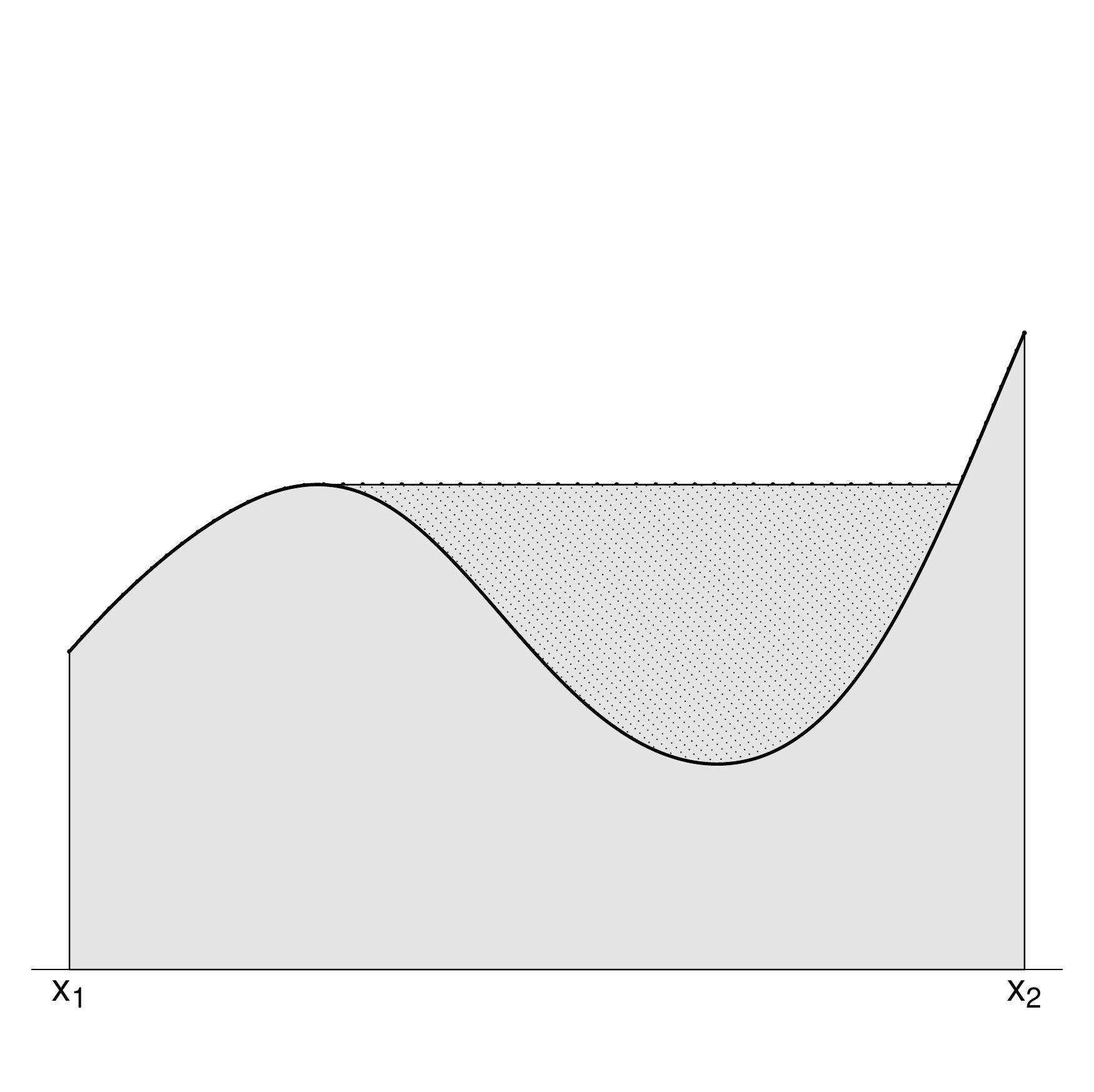}
  }
  \caption{The left panel displays a set of points in $\Real^2$, of which
 two pairs, $(x_1, x_2)$ and $(x_3, x_4)$, are highlighted by drawing the
 segment joining them. The right plot displays the section of $\hat{f}(x)$
 along the segment which joins $x_1$ and $x_2$ (smooth curve delimiting the
 light-shaded area) and the associated auxiliary non-decreasing function (with
 filled area highlighted).} 
  \label{fig:pairs}
\end{figure}

This broad idea of tolerance must be given a specific form to be operational.
\citet{Menardi_Azzalini2012} adopt a criterion which for simplicity we only
describe informally, and refer the reader to their paper for full
specification.  When a valley is detected, we introduce an auxiliary function
derived from the original one by increasing it by the minimum amount required
to fill the valley; one can think that water is poured into the valley until
it starts to overflow.  To visualize this process, consider
Figure~\ref{fig:pairs} where, in the first panel, a sample of points in
$\Real^2$ is depicted and the segments joining two pairs of them, $(x_1, x_2)$
and $(x_3, x_4)$, are highlighted.  The second panel displays the section of
$\hat{f}(x)$ along the segment joining $x_1$ and $x_2$, represented by the
smooth curve, and the auxiliary function which fills the valley. The
amplitute of the valley in this case is quantified by
\[
 R = \frac{\mbox{integral of the dark-shaded area}}%
    {\mbox{integral of the dark-shaded area $+$ integral of the light-shaded area}}
   \in[0,1).
\]\label{eq:area}
If $R<\lambda$, for a given tolerance parameter $\lambda\in(0,1)$, $x_1$ and
$x_2$ are considered connected points, and an edge is set in the connection
graph.  For the pair $(x_3, x_4)$ in Figure~\ref{fig:pairs}, the section of 
$\hat{f}(x)$ along the segment joining them is concave (plot
not shown); hence in this case $R=0$ and the points are declared to be
connected.

Once the connection of all pairs of sample values has been examined, the rest
of the process is carried out exactly as in the previous section. Since 
now we are always working in a one-dimensional world, 
higher values of $d$ can be tackled. However, for large $d$ we are facing
another problem:  the degrade of performance of nonparametric
density estimates, known as `course of dimensionality'.
On the other hand,  it can be argued that for the clustering problem we 
 need to identify only the main features of a distribution, especially
its modes, not the fine details. This consideration indicates than the
method can be considered also for a broader set of cases than those where
the density is the focus of interest. See \citet{Menardi_Azzalini2012} for
a more extended discussion of this issue.

\section[The R package pdfCluster]{The \proglang{R} package \pkg{pdfCluster}}

\subsection{Package overview}
The \proglang{R} package \pkg{pdfCluster} performs, as its main task, cluster analysis by implementing the methodology described in the previous sections.
Furthermore, two other tasks are accomplished: density estimation and clustering diagnostics.  

Each of these tasks is associated with a set of functions and methods, and results in an object of a specific class, as summarized in Table~\ref{tab:summary_pack}.  The package is built by making use of classes and
 methods of the \proglang{S4} system.

\begin{table}
\begin{tabular}{p{.25\textwidth}p{.25\textwidth}p{.22\textwidth}p{.17\textwidth}}
\hline & clustering & density estimation & diagnostics \\ \hline
\proglang{S4} class & 
\code{pdfCluster-class} & \code{kepdf-class} & \code{dbs-class}\\
 &&&\\
functions to produce \par the class & 
\code{pdfCluster}\par \code{pdfClassification} & \code{kepdf} & \code{dbs}\\
 &&&\\
related methods & 
\code{pdfCluster} \par\code{plot} \par\code{show} \par\code{summary} & \code{plot} \par \code{show} \par\code{summary} & \code{dbs} \par\code{plot} \par\code{show} \par\code{summary}\\ 
others &
& \code{h.norm}\par\code{hprop2f} & \code{adj.rand.index}\\
\hline
\end{tabular}
\caption{Summary of the \pkg{pdfCluster} package. Each of the three main tasks of the 
package is associated with a set of functions and methods, and results in an object of
a specific \proglang{S4} class.}\label{tab:summary_pack}
\end{table}

To ease presentation, an overview of the package is first provided, aiming at an
introductory usage of its features. The next section is devoted to a more in-depth  
examination of computational and technical details. 

\subsubsection*{Clustering via nonparametric density estimation} 
The package is built around the main routine \code{pdfCluster}, which actually performs
 the clustering process. \code{pdfCluster} is defined as a generic function and 
dispatches different methods depending on the class of its first argument. 
For a simple use of this function, the user is only required to provide the data 
\code{x} to be clustered, in the form of a vector, \code{matrix}, or \code{data.frame}
 of \code{numeric} elements. 
A further method dispatched by the \code{pdfCluster} generic function applies to objects
 of \code{pdfCluster-class} itself. This last option will be discussed later.

Further arguments include \code{graphtype}, which defines the procedure to identify the
 connected components of the density level sets. The elementary case $d=1$
is handled by the \code{"unidimensional"} option. When data are multidimensional, instead, this argument 
may be set to \code{"delaunay"}  or \code{"pairs"}, to run the procedures, 
as described in Section \ref{sec:delaunay} and \ref{sec:pairs}, respectively. If 
not specified, the latter option is selected when the data dimension is greater than $6$. 
When \code{"pairs"} is selected, explicitly or implicitly, the user may  wish to 
specify the parameter \code{lambda}, which defines the tolerance threshold to claim
 the presence of valleys in the section of $\hat{f}(x)$ between pairs of points. 
Default value is \code{lambda=0.10}.


After that the connected components associated with the density level sets are identified, 
\code{pdfCluster} builds the cluster tree and detects the cluster cores. An internal 
call to function \code{pdfClassification} follows by default, to carry on the 
second phase of the clustering procedure, that is allocation of the lower density 
data points not belonging to the cluster cores. The user can regulate the process 
of classification by setting some optional parameters to be passed to 
\code{pdfClassification}. Details are discussed below.  

Further optional arguments may given to \code{pdfCluster} in order to regulate 
density estimation. These arguments are passed internally to function \code{kepdf}, 
which is described below.
      
The results of the clustering process are encapsulated in an object of 
\code{pdfCluster-class},  whose slots include, among others, a list \code{nc} 
providing details about the connected components identified for each considered 
section of the density estimate, a vector \code{cluster.cores}
 defining the cluster cores membership, and an object of class \code{dendrogram}, providing information 
about the cluster tree, the number \code{noc} of detected groups. Furthermore, when the classification 
procedure is carried on, the slot \code{stages} is  a list with elements corresponding to the data allocation 
to groups at the different stages of the classification procedure and \code{clusters} reports the final 
group labels.
 
Methods to \code{show}, to provide a \code{summary} and to \code{plot} objects of 
\code{pdfCluster-class} are available. Four types of plot are selectable, by setting 
the argument \code{which}. Argument \code{which=1} plots the mode function; \code{which=2} 
plots the cluster tree; a scatterplot of the data or of pairs of selected coordinates 
reporting the label group is provided when \code{which=3} and \code{which=4} plots the 
density-based silhouette information as will be described below. Multiple choices are possible.

\subsubsection*{Density estimation} 
Density estimation is performed by the kernel method througout the \code{kepdf} function. 
Estimates are computed by a product estimator of the form: 
\[
\hat{f}(y)= \sum_{i=1}^n \frac{1}{n h_{i,j} \cdots h_{i,d}} \prod_{j=1}^d K\left(\frac{y_{j} - x_{i,j}}{h_{i,j}}\right).
\]
The kernel function $K$ is an argument of function \code{kepdf} and can either be a Gaussian density 
(if \code{kernel = "gaussian"}) or a $t_\nu$ density, with $\nu = 7$ degrees of freedom (when \code{kernel = "t7"}).
The uncommon option of selecting a $t$ distribution is motivated by computational reasons, 
as will be clarified in Section \ref{sec:details}.
 
The user may choose to estimate density with a fixed or an adaptive bandwidth 
$h_{i} = (h_{i,1} \cdots h_{i,d})'$, by setting the argument \code{bwtype} accordingly. 
Leaving the argument unspecified entails the use of a fixed bandwidth estimator. 
When \code{bwtype="fixed"}, that is $h_{i} = h$, a constant smoothing 
vector is used for all the observations $x_i$. Default values are set as asymptotically 
optimal for a multivariate Normal distribution \citet[see, e.g.,][page 32]{Bowman_Azzalini97}. 
Alternatively, \code{bwtype="adaptive"}, which corresponds to specify a vector of bandwidths $h_i$ 
for each observation $x_i$. Default values are selected according to the approach described by
\citet[Section 5.3.1]{Silverman86}, implemented in the package through the function \code{hprop2f}. 

Results of the application of the \code{kepdf} function are encapsulated in objects of \code{kepdf-class}, whose slots include the \code{estimate} at the evaluation points and the parameters used to obtain that estimate.

Methods which \code{show}, provide a \code{summary}, and \code{plot} objects of 
\code{kepdf-class} are also available. When the density estimate is based on two or 
higher dimensional data, these functions make use of functions \code{contour}, 
\code{image} and \code{persp}, depending on how the argument \code{method} is set. 
When $d>2$, the pairwise marginal estimated densities are plotted for all pairs of coordinates, or for a subset of coordinates specified by the user via the argument \code{indcol}.

\subsubsection*{Diagnostics of clustering} 
As a third feature, the package provides diagnostics of the clustering outcome. 
Density-based silhouette is computed by the generic function \code{dbs}, which 
dispatches two methods. 
One method applies to objects of \code{pdfCluster-class} directly; a second 
method is thought to compute the density-based silhouette information on partitions 
produced by a possibly different density-based clustering technique. The latter method 
applies applies to two arguments: the matrix of clustered data and a numeric vector of 
cluster labels.

Computation of the density-based silhouette requires the density function to be estimated, 
conditional to the group membership. Hence, further arguments of function \code{kepdf} 
can be given to \code{dbs} to set parameters of density estimation. 
Moreover, some \code{prior} probability may be specified for each cluster.  

Results of the application of function \code{dbs} are provided in objects of 
\code{dbs-class}. An \proglang{S4} method for plotting objects of \code{dbs-class} 
is available: data are partitioned into the clusters, sorted in a decreasing order 
with respect to their \emph{dbs} value and  displayed on a bar graph. 

As a further diagnostic tool, the package provides the function \code{adj.rand.index}
which evaluates the agreement between two partitions, through 
the adjusted Rand criterion \citep{Hubert_Arabie85}. 

\subsection{Further details}\label{sec:details}

\begin{itemize}
\item As already mentioned, \code{pdfCluster} automatically selects the procedure 
to be used for detecting connected components of the density level sets, depending on 
the data dimensionality. While the user is allowed to change this choice by setting 
argument \code{graphtype}, we warn against setting the argument \code{graphtype="delaunay"}
for large dimensions. The number of operations required to compute the Delaunay 
triangulation grows with $n^{\lfloor d/2 \rfloor}$ while the computational complexity due
 to run the pairwise connection criterion grows quadratically with the sample size. 
Hence, at the present state of computing resourses, running the Delaunay triangulation 
when $d> 6$ appears hardly feasible for values of $n$ greater than about $200$. 
Instead, data with any dimensionality may be handled by the pairwise connection criterion, although 
the computational speed slows down for very large $n$.   
\item The higher computational efficiency of the pairwise connection criterion 
is paid for by the need of setting the tolerance threshold $\lambda$.
According to our experience, the default value of \code{lambda=0.10} is usually a 
sensible choice for moderate to high dimension while a lower \code{lambda} is sufficient 
in low dimensions, when the density estimate is more accurate. A larger value can be 
useful when the procedure detects a number of small spurious groups, because this choice 
results in aggregating clusters. 
\item As running the procedure several times with different choices of \code{lambda} 
may be time consuming, the package allows for a more efficient route, implemented by an 
additional method dispatched by function \code{pdfCluster}. Once that an object of 
\code{pdfCluster-class} is created by function \code{pdfCluster} with argument 
\code{graphtype= "pairs"}, \code{pdfCluster} can be called again by setting the 
same \code{pdfCluster-class} object as a first argument \code{x} and a different 
value of \code{lambda}: slot \code{graph} of the \code{pdfCluster-class} object 
contains the amplitude of the valleys detected by the evaluation of the density 
along the segments connecting observations. Then, the pairwise evaluation does not 
need to be run again to check results for different values of \code{lambda} and the 
procedure speeds up considerably. An example will be illustrated in the next section. 
\item Both the Delaunay and the pairwise connection criterion to build a graph on the 
observed data are implemented by some specifically designed foreign functions. 
In the former case, the \code{delaunayn} function in package \pkg{geometry} 
\citep{Geometry} is the \proglang{R} interface to the Quickhull algorithm. 
Pairwise connection is, instead, implemented in the \proglang{C} language.
\item After building the selected connection network among the observations, 
\code{pdfCluster} determines the level set $S(c)$ for a grid of values of $p_c$; 
the length of such grid may be set through the \code{n.grid} argument. For each 
value of $p_c$ the identification of connected components of $S(c)$ is carried out 
by means of a \proglang{C} function borrowed by the \proglang{R} package \pkg{spdep} 
\citep{Spdep}, which implements a depth first search algorithm. 
\item The procedure to allocate the low density data to the cluster cores is 
block-sequential and the user is allowed to select the number \code{n.stage} of 
such blocks as an optional parameter of \code{pdfCluster}, to be passed internally 
to \code{pdfClassification} (default value is \code{n.stage = 5}). When this argument is set to $0$, the clustering process 
stops when the cluster cores are identified. Otherwise, further arguments can be passed from \code{pdfCluster} to \code{pdfClassification}. Among them, \code{se} takes \code{"logical"} values, and is set to \code{TRUE} to account for the standard error of the log-likelihood ratios in Equation \ref{pdf-ratio}. 
Argument \code{hcores} declares if the densities in (\ref{pdf-ratio}) have to be 
estimated by selecting the same bandwidths as the ones used 
to form the cluster cores. Default 
value is set to \code{FALSE}, in which case a vector of bandwidth specific for the clusters is used.
\item \code{pdfCluster} makes an internal call to function \code{kepdf} both to estimate the density underlying data and to build the connection network when the pairwise connection criterion is selected. By default, a kernel density estimation with fixed kernel is built, with vector of smoothing parameters set to the one asymptotically optimal under the 
assumption of multivariate normality. Although arguably sub-optimal, this choice produces sensible results in most applications. When dimensionality of data is low-to-moderate, it is often advantageous to shrink the smoothing parameter slightly towards zero; we adopt a shrinkage factor \code{hmult=3/4} when $d \leq 6$, 
as recommended by \citet{AT}, but the default value may be optionally modified by the user. For higher-dimensional data, instead, we suggest the use of a kernel estimator with adaptive bandwidth, which can be obtained by setting argument \code{bwtype} to \code{"adaptive"}. 
\item Function \code{kepdf} represents the \proglang{R} interface of two \proglang{C} routines which allow to speed computations. Each of these routines is designed to perform kernel density estimation with a specific kernel function. It is worth to remind that, when connected sets are identified by the pairwise connection criterion, computation of the $R$ measure in Equation~\ref{eq:area} requires the density function to be evaluated along 
the segments joining each pair of observations. In practice, a grid of \code{grid.pairs} points is considered for each segment, so that the number of operation required grows with $n^2$\code{grid.pairs} . \\ 
When sample size is very large, any saving in the arithmetic computations of the kernel can make a noticeable difference. In particular, the use of a 
Gaussian kernel requires a call to the exponential function at each evaluation, which is computationally 
much more expensive than sum, multiplication and power function. This explains the non-standard 
option to select a $t_\nu$ kernel, with $\nu=7$ degrees of freedom. The relatively more 
critical computation involves an $8$th degree power, which can be coded efficiently and still the kernel has unbounded support, 
which is more appropriate for the classification stage than alternatives like bisquare or similar kernels.  
The use of this option is then suggested when the sample size is huge.     
\item To compute \code{dbs}, it is possible to specify some prior probability of each cluster. To this end, 
recall that clusters are labelled according to the maximal value of the density, so that label $1$ refers to the 
cluster with overall maximal density. The choice depends on the prior knowledge about the composition of the clusters and a lack of information would imply the choice of
a uniform distribution over the groups. However, information derived from the detected partition can also be
used. When diagnostics are computed on an object of \code{pdfCluster-class}, prior probabilities can be chosen as proportional to the cardinalities of the cluster cores. This is also the default choice. In a model-based clustering, instead, the mixing proportions could be a natural choice.
\end{itemize}
\section{Some illustrative examples}
\subsection{Quantitative variables}
The \code{wine} data set was introduced by \citet{Wine}. It originally included the results of 27 chemical measurements on 178 wines grown
in the same region in Italy but derived from three different
cultivars: Barolo, Grignolino and Barbera. The \pkg{pdfCluster} package provides a selection of 13 variables. The data set is here used to illustrate the main features of the package. 

\begin{figure}
  \centerline{
   \includegraphics[width=0.55\hsize]{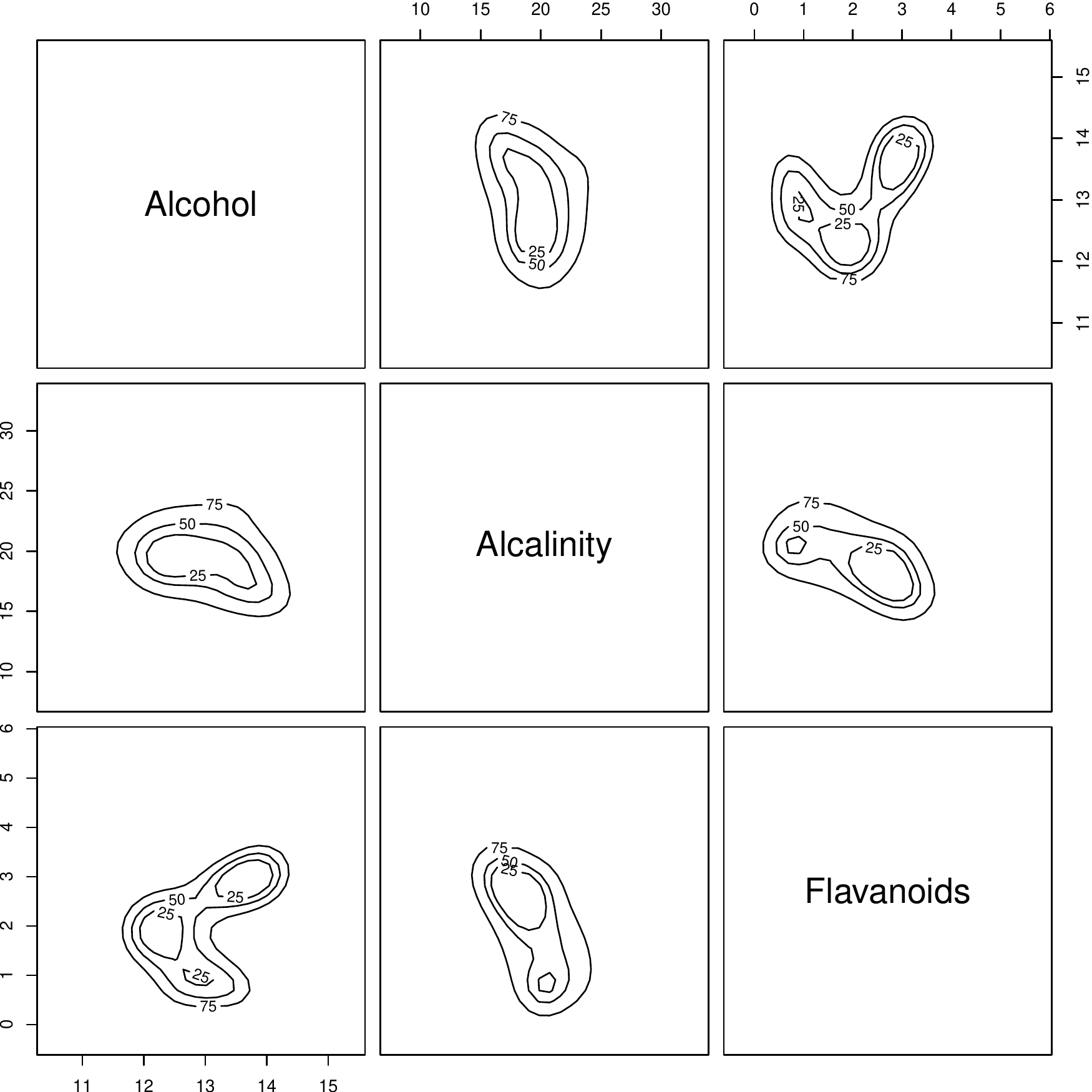}
  }
  \caption{Plot of the pairwise marginal density estimates of three variables 
of \emph{wine} data, as given by function \code{kepdf}.} 
  \label{fig:kepdf_plot}
\end{figure}

As a first simple example, let us suppose to have some knowledge about which  variables are relevant to the aim of reconstructing the cultivar of origin of each wine. We then perform cluster analysis on a small subset of variables.
\begin{CodeChunk}
\begin{CodeInput}
R> library("pdfCluster")
R> data("wine")
R> winesub <- wine[, c(2,5,8)]
\end{CodeInput}
\end{CodeChunk}
As the number of considered variables is very small, visual exploration of the density estimate of the data may already give us some indication about the clustering structure.
\begin{CodeChunk}
\begin{CodeInput}
R> pdf <- kepdf(winesub)
R> plot(pdf, text.diag.panel= names(winesub))
\end{CodeInput}
\end{CodeChunk}
The resulting plot is displayed in Figure~\ref{fig:kepdf_plot}. A three-cluster structure is clearly evident from the marginal density estimate of the variables "Alcohol" and "Flavanoids".
The main content of the \code{kepdf-class} object \code{pdf} may be printed by the associated \code{show} method.  
\begin{CodeChunk}
\begin{CodeInput}
R> pdf
\end{CodeInput}
\begin{CodeOutput}
An S4 object of class "kepdf"

Call: kepdf(x = winesub)

Kernel:  
[1] "gaussian"

Estimator type: fixed bandwidth 

Diagonal elements of the smothing matrix:  0.3750856 1.542968 0.4614995 

Density estimate at evaluation points:  
  [1] 0.015211471 0.001994922 0.009822658 0.010526400 0.009014892 0.013104296
  [7] 0.005910667 0.013900582 ...
\end{CodeOutput}
\end{CodeChunk}   
Clustering is performed by a call to \code{pdfCluster}. Note that its usage does not require a preliminary call to \code{kepdf}. A \code{summary} of the resulting object provides the cardinalities of both the cluster cores and the clusters, along with the main structure of the cluster tree.
\begin{CodeChunk}
\begin{CodeInput}
R> cl.winesub <- pdfCluster(winesub)
R> summary(cl.winesub)
\end{CodeInput}
\begin{CodeOutput}
An S4 object of class "pdfCluster"

Initial groupings: 
 label:    1   2   3  NA 
 count:   29  13  17 119 

Final groupings: 
 label:   1  2  3 
 count:  62 63 53 

Groups tree:
--[dendrogram w/ 1 branches and 3 members at h = 1]
  `--[dendrogram w/ 2 branches and 3 members at h = 0.36]
     |--leaf "1 " 
     `--[dendrogram w/ 2 branches and 2 members at h = 0.32]
        |--leaf "2 " (h= 0.04  )
        `--leaf "3 " (h= 0.06  )
\end{CodeOutput}
\end{CodeChunk} 

The object may be further inspected by accessing its slots. Slot \code{graph}, 
for instance, discloses the procedure used to find the connected components 
associated to the level set:
\begin{CodeChunk}
\begin{CodeInput}
R> cl.winesub@graph
\end{CodeInput}
\begin{verbatim}
$type
[1] "delaunay"
\end{verbatim}
\end{CodeChunk} 
The user may be also interested about details regarding the estimated density, 
available through the slot \code{pdf}. 
\begin{CodeChunk}
\begin{CodeInput}
R> cl.winesub@pdf
\end{CodeInput}
\begin{verbatim}
$kernel
[1] "gaussian"

$bwtype
[1] "fixed"

$par
$par$h
   Alcohol Alcalinity Flavanoids 
 0.2813142  1.1572259  0.3461246 

$par$hx
NULL

$estimate
  [1] 0.021153490 0.003723019 0.009561598 0.013346244 0.011821547 0.017818041
  [7] 0.006527976 0.017082718 ...
\end{verbatim}
\end{CodeChunk} 

Note that the vector of smoothing parameters used to estimate density, during 
the process of clustering, differs from the one produced by the previous call 
to function \code{kepdf}, whose default value is asymptotically optimal 
for Normal data, as given by function \code{h.norm()}. As already mentioned, 
when low-dimensional data are clustered (in this example $d=3$), this vector 
is multiplied by a shrinkage factor of $3/4$ by default.  
  
Additional information about the detected partition may be further visualized 
through a call to the associated \code{plot} methods. If argument \code{which} 
is not selected, the four available types of plot are displayed one at a time. 

\begin{figure}
\begin{tabular}{p{.5\textwidth}p{.5\textwidth}}
    \includegraphics[width=\hsize]{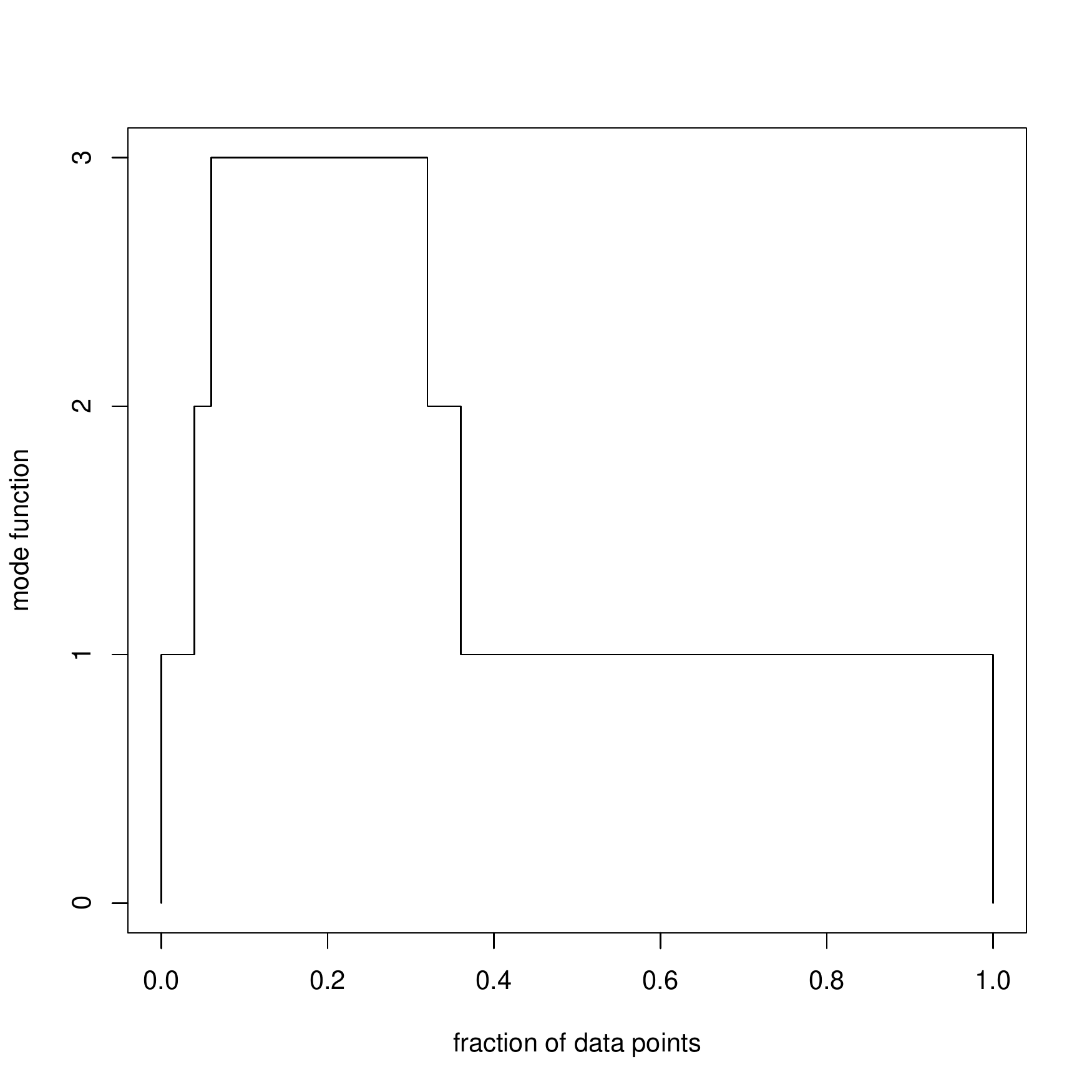}
    \hfill&
    \includegraphics[width=\hsize]{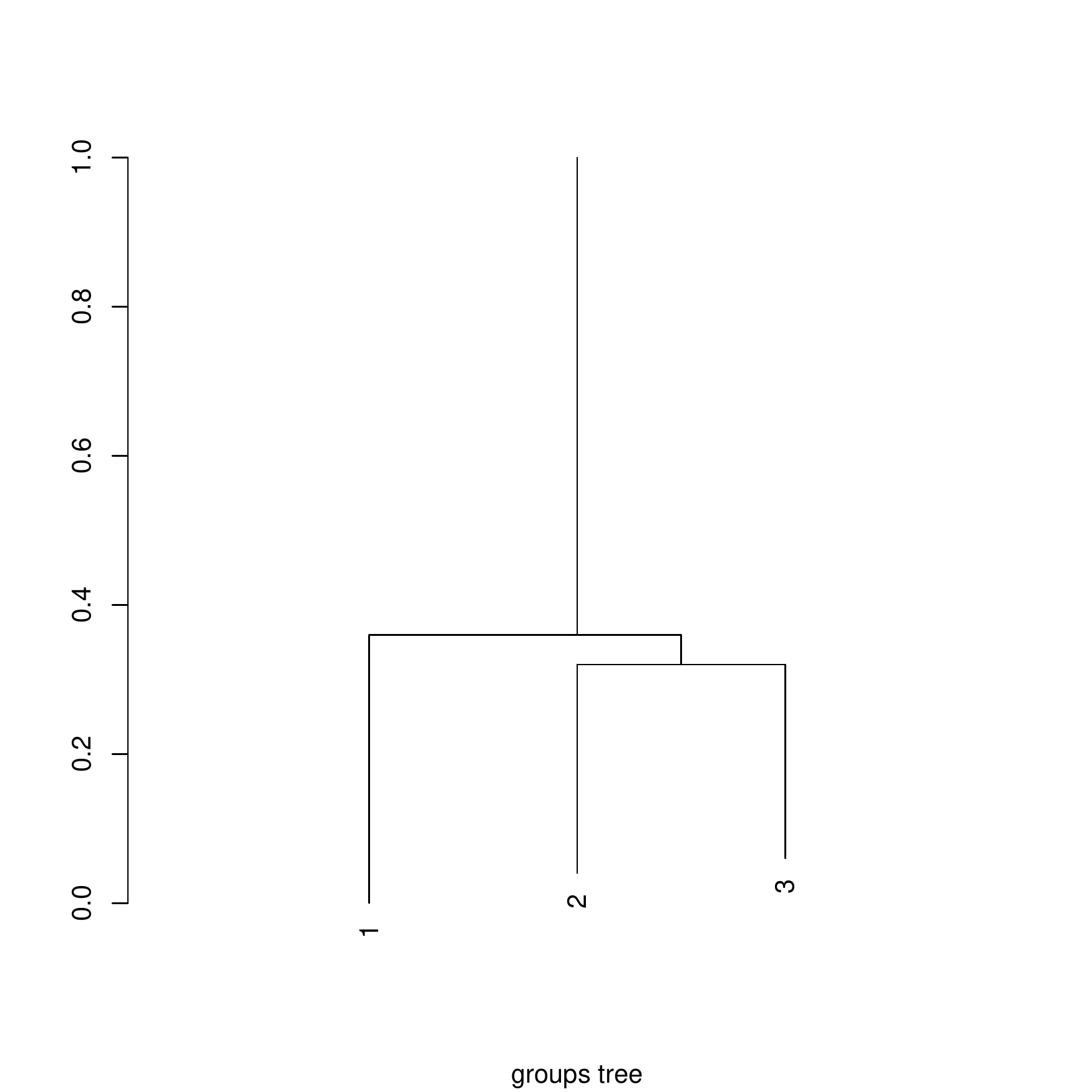}\\     
\includegraphics[width=1.11\hsize]{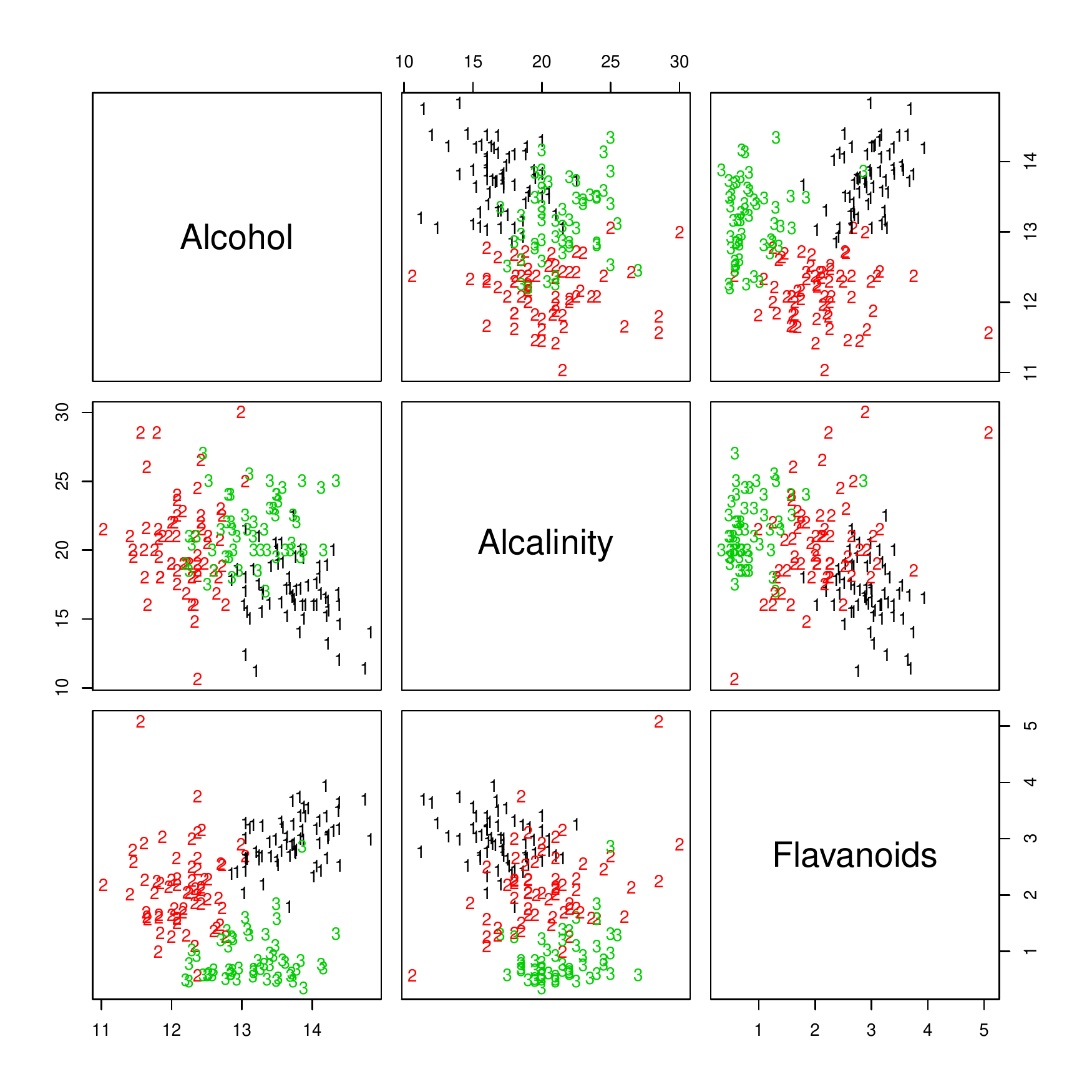}&
\hspace{-1cm}
\includegraphics[width=1.11\hsize]{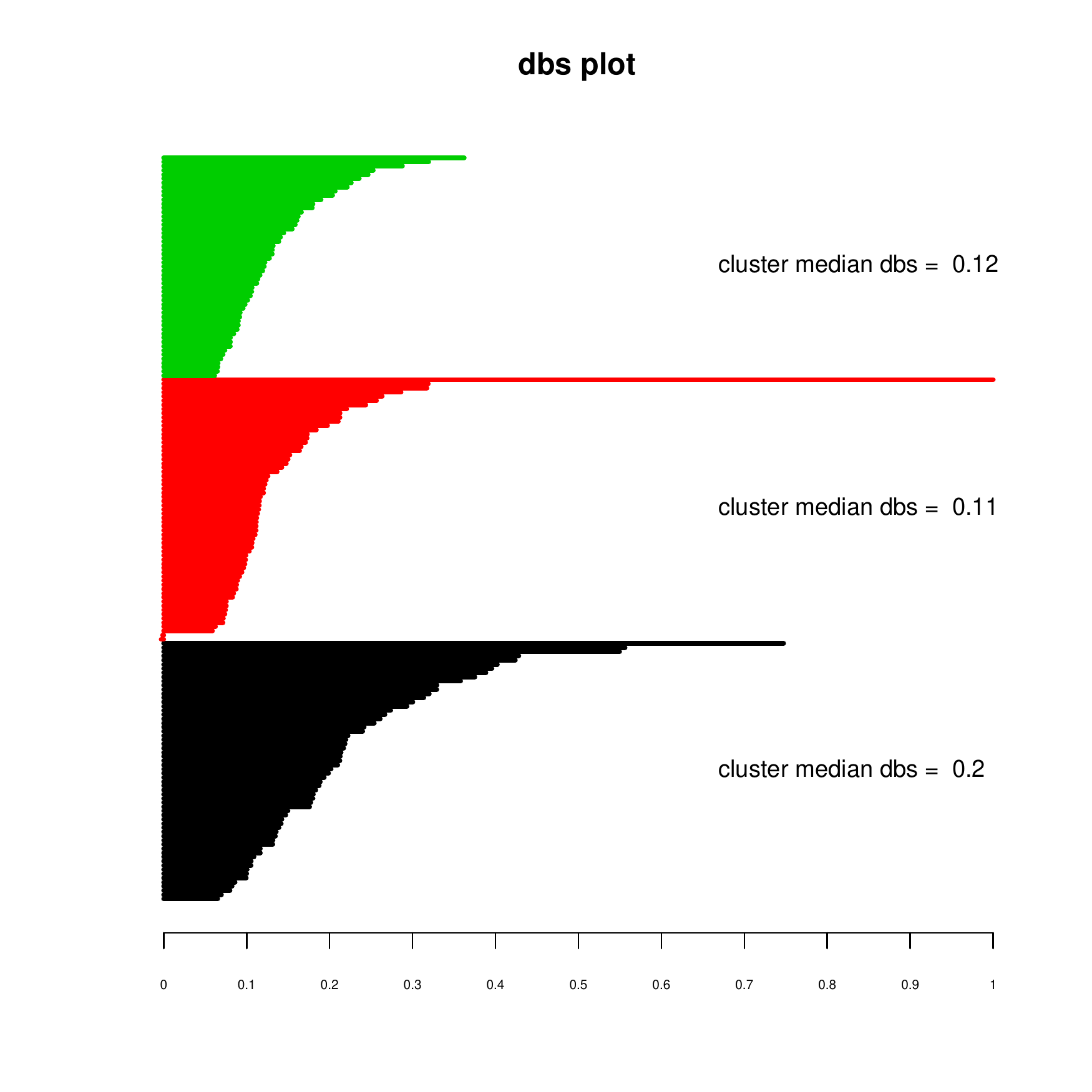}
  \end{tabular}
  \caption{Result of the application of plot methods on objects of \code{pdfCluster-class}. From the left: the mode function, the cluster tree, the pairwise scatterplot and the \emph{dbs} plot.} 
  \label{fig:pdfCluster_plot}
\end{figure}

\begin{CodeChunk}
\begin{CodeInput}
R> plot(cl.winesub)
\end{CodeInput}
The resulting plots are reported in Figure~\ref{fig:pdfCluster_plot}. In particular, the diagnostic plot presents values of the \code{dbs} appreciably larger than zero for almost all the observations throughout the three clusters, suggesting the soundness of the detected partition. This is confirmed by the cross-classification frequencies of the the obtained clusters and the actual cultivar of origin of the wines (indicated in the first column of the \code{wine} data).
\begin{CodeInput}
R> table(wine[,1], cl.winesub@clusters)
\end{CodeInput}
\begin{CodeOutput}
              1  2  3
  Barolo     58  1  0
  Grignolino  4 62  5
  Barbera     0  0 48
\end{CodeOutput}
\end{CodeChunk}
  
Consider now the entire data set \code{wine} (we only remove the true label class of wines).  
\begin{CodeChunk} 
\begin{CodeInput}
R> wineall <- wine [, -1]
\end{CodeInput}
\end{CodeChunk} 
When high-dimensional data are clustered, some caution should be used to deal with 
the curse of dimensionality and the increased variability of the density estimate. 
\citet{Menardi_Azzalini2012} suggest either to allow for different amount of 
smoothing by using an adaptive bandwidth, or to oversmooth the density:
\begin{CodeChunk}
\begin{CodeInput}
R> cl.wineall <-pdfCluster(wineall, bwtype="adaptive")
R> summary(cl.wineall)
\end{CodeInput}
\begin{CodeOutput}
An S4 object of class "pdfCluster"

Initial groupings: 
 label:    1   2   3   4   5   6  NA 
 count:    5   2  10   4   5   3 149 

Final groupings: 
 label:   1  2  3  4  5  6 
 count:  35 32 36 40 20 15 

Groups tree:
--[dendrogram w/ 1 branches and 6 members at h = 1]
  `--[dendrogram w/ 2 branches and 6 members at h = 0.62]
     |--[dendrogram w/ 1 branches and 4 members at h = 0.32]
     |  `--[dendrogram w/ 3 branches and 4 members at h = 0.28]
     |     |--[dendrogram w/ 2 branches and 2 members at h = 0.04]
     |     |  |--leaf "2 " (h= 0.02  )
     |     |  `--leaf "1 " 
     |     |--leaf "3 " (h= 0.12  )
     |     `--leaf "4 " (h= 0.12  )
     `--[dendrogram w/ 2 branches and 2 members at h = 0.32]
        |--leaf "5 " (h= 0.14  )
        `--leaf "6 " (h= 0.18  )
\end{CodeOutput}
\end{CodeChunk} 

Note that six groups have now been produced. The identification of more than three 
clusters, when all the thirteen variables of the \code{wine} data are used, is 
consistent with results of the application of other clustering methods 
\citep[see, e.g., ][]{Mcnicholas_Murphy08}.
However, their pairwise aggregation as
$\{1, 2\}, \{3, 6\}, \{4, 5\}$ essentially leads to the three actual cultivars, with 
a few misclassified points left: 
\begin{CodeChunk}
\begin{CodeInput}
R> table(wine[,1], cl.wineall@clusters) 
\end{CodeInput}
\begin{CodeOutput}    
              1  2  3  4  5  6
  Barolo     32 27  0  0  0  0
  Grignolino  3  5  0 40 19  4
  Barbera     0  0 36  0  1 11
\end{CodeOutput}
\end{CodeChunk} 
A more accurate classification may be obtained by additionally employing
 oversmoothing the density: 
\begin{CodeChunk}
\begin{CodeInput}
R> cl.wineall.os <-pdfCluster(wineall, bwtype="adaptive", hmult=1.2)
R> table(wine[,1], cl.wineall.os@clusters)
\end{CodeInput}
\begin{CodeOutput}               
              1  2  3
  Barolo     59  0  0
  Grignolino  6  4 61
  Barbera     0 47  1
\end{CodeOutput}
\end{CodeChunk} 

A similar effect may be caused by relaxing the condition to set connections among points, 
while building the pairwise connection graph among the observations.  
In the current example, data dimensionality equal to $13$ has entitled the use of 
the pairwise connection criterion, as may be seen from:
\begin{CodeChunk}
\begin{CodeInput}
R> cl.wineall@graph[1:2]
\end{CodeInput}
%/begin{verbatim}
\begin{CodeOutput}
$type
[1] "pairs"

$lambda
[1] 0.1
\end{CodeOutput}
%\end{verbatim}
\end{CodeChunk} 

In addition to the criterion adopted to build the network connection, the slot 
\code{graph} reports the tolerance threshold \code{lambda}. 
The third, not displayed, element of the slot contains the amplitude of the 
valleys detected by the evaluation of the density along the segments 
connecting all the pairs of observations. 
A larger value of $\lambda$ typically results in aggregation of clusters, 
and can be obtained either by  re-running the whole procedure:
\begin{CodeChunk}
\begin{CodeInput}
R> cl.wineall.l0.2 <- pdfCluster(wineall, bwtype = "adaptive", lambda = 0.2)
\end{CodeInput} 
\end{CodeChunk} 
or by applying the method which the \code{pdfCluster} function dispatches to \code{pdfCluster-class} objects: 
\begin{CodeChunk}
\begin{CodeInput}
R> cl.wineall.l0.2 <- pdfCluster(cl.wineall, lambda = 0.2)
\end{CodeInput}
\end{CodeChunk}  
The latter choice does not re-run pairwise evaluation and it exploits, instead, 
computations saved in the slot \code{graph} of the \code{pdfCluster-class} object. 
This speeds up the clustering procedure considerably, especially when the sample 
size is large. 

\subsection{Mixed variables}

As they stand, density-based clustering methods may be applied to continuous data only. 
However, we illustrate here how this assumption may be circumvented.

Consider the data set \code{plantTraits} from package \pkg{cluster} \citep{Cluster}. It 
describes 136 plant species according to 31 morphological and 
reproductive attributes. 
\begin{CodeChunk}
\begin{CodeInput}
R> library("cluster")
R> data("plantTraits")
\end{CodeInput}
\end{CodeChunk}
In order to use all available information to cluster the data, a reasonable 
procedure seems to us as follows: first, a dissimilarity matrix among the points 
is created, using criteria commonly employed in classical distance-based methods; 
next, from this matrix, a configuration of points in the $d'$-dimensional 
Euclidean space is sought by means of multidimensional scaling, for some given $d'$.
In this way, $d'$ numerical coordinates are obtained, to be passed to the clustering
procedure.

It is inevitable that any recoding procedure of this sort involves some 
arbitrarieness and the one just described is no exception. However, of the two 
steps involved, the first one is exactly that considered by classical hierarchical clustering techniques. The second step requires a few additional choices, especially the 
number $d'$ of principal coordinates. A detailed exploration of these aspects is 
definitely beyond the scope of this paper and will be pursued elsewhere. 

For the present illustrative purposes, we refer to the example reported in the 
documentation of the \pkg{cluster} package, and we choose the Gower coefficient to measure 
the dissimilarity among the points, implemented in \proglang{R} by function 
\code{daisy}. The categorical variables are distinguished in ordered, binary 
asymmetric and symmetric, and handled accordingly \citet[see][Chapter 1]{Kauf:Rous:1990}.  
\begin{CodeChunk}
\begin{CodeInput}
R> example(plantTraits)
...
plntTr> dai.b <- daisy(plantTraits,
plntTr+		type = list(ordratio=4:11, symm=12:13, asymm=14:31)) 
\end{CodeInput} 
\end{CodeChunk}
We then make use of classical multidimensional scaling \citep{Gower66}, which is
 computed by function \code{cmdscale}, and we take the simple option of setting $d'=6$ 
as in \citet{Cluster}.
\begin{CodeChunk}
\begin{CodeInput}
plntTr> cmdsdai.b <-cmdscale(dai.b, k=6) 
\end{CodeInput} 
\end{CodeChunk}
With the new set of data, we may run the clustering procedure. 
\begin{CodeChunk}
\begin{CodeInput}
R> cl.plntTr <-pdfCluster(cmdsdai.b)
R> summary(cl.plntTr)
\end{CodeInput}  
\begin{CodeOutput}
An S4 object of class "pdfCluster"

Initial groupings: 
 label:    1   2  NA 
 count:   23   2 111 

Final groupings: 
 label:    1   2 
 count:  128   8 

Groups tree:
--[dendrogram w/ 1 branches and 2 members at h = 1]
  `--[dendrogram w/ 2 branches and 2 members at h = 0.18]
     |--leaf "1 " 
     `--leaf "2 " (h= 0.16  )
\end{CodeOutput}  
\end{CodeChunk}
The outcome indicates the presence of two clusters. However, the number of 
data points assigned to one of the two clusters is very small, and the associated 
cluster core is formed by two observations only. In these circumstances, selecting a
global bandwidth to classify the lower density data seems to be more appropriate 
than using cluster specific bandwidths. This can be pursued with the following commands: 
\begin{CodeChunk}
\begin{CodeInput}
R> cl.plntTr.hc <-pdfCluster(cmdsdai.b, cores=TRUE)
\end{CodeInput}  
\end{CodeChunk}
While the \code{plantTraits} data set does not include information about a true 
label class of the cases, it is interesting to note that the two identified groups
roughly correspond to the aggregation of the clusters $\{1,3,5,6\}$ and $\{2,4\}$
identified by running a distance-based method and then cutting the dendrogram 
at six clusters, as suggested by \citet{Cluster}. 
\begin{CodeChunk}
\begin{CodeInput}
plntTr> agn.trts <- agnes(dai.b, method="ward")
...
plntTr> cutree6 <- cutree(agn.trts, k=6)
R> table(cutree6, cl.plntTr.hc@clusters)
\end{CodeInput}  
\begin{CodeOutput}
cutree6  1  2
      1 10  0
      2 11 20
      3 21  0
      4  4 15
      5 18  1
      6 35  1
\end{CodeOutput}  
\end{CodeChunk}
Note that selecting six principal coordinates places us at the threshold we defined 
to choose both the type procedure to find the connected level sets and the way of 
smoothing the density function. Since the threshold $d=6$ is merely indicative, in these
circumstances it makes sense to check results deriving from a different setting of 
the arguments as, for example
\begin{CodeChunk}
\begin{CodeInput}
R> cl.plntTr.hc.newset <-pdfCluster(cmdsdai.b, cores=TRUE, graphtype="pairs",
				     bwtype="adaptive")
\end{CodeInput} 
\end{CodeChunk} 
A variant procedure to handle mixed data would recode the categorical variables
 only, and run \code{pdfCluster} on the set of data obtained by merging 
the principal coordinates so constructed and the original quantitative variables.  
\bibliography{biblio} 

\begin{thebibliography}{26}
\newcommand{\enquote}[1]{``#1''}
\providecommand{\natexlab}[1]{#1}
\providecommand{\url}[1]{\texttt{#1}}
\providecommand{\urlprefix}{URL }
\expandafter\ifx\csname urlstyle\endcsname\relax
  \providecommand{\doi}[1]{doi:\discretionary{}{}{}#1}\else
  \providecommand{\doi}{doi:\discretionary{}{}{}\begingroup
  \urlstyle{rm}\Url}\fi
\providecommand{\eprint}[2][]{\url{#2}}

\bibitem[{Azzalini \emph{et~al.}(2012)Azzalini, Menardi, and
  Rosolin}]{PdfClusterpkg}
Azzalini A, Menardi G, Rosolin T (2012).
\newblock \emph{\proglang{R} Package \pkg{pdfCluster}: Cluster Analysis Via
  Nonparametric Density Estimation (Version 1.0-0)}.
\newblock Universit\`a di Padova, Italia.
\newblock
  \urlprefix\url{http://cran.r-project.org/web/packages/pdfCluster/index.html}.

\bibitem[{Azzalini and Torelli(2007)}]{AT}
Azzalini A, Torelli N (2007).
\newblock \enquote{Clustering Via Nonparametric Density Estimation.}
\newblock \emph{Statistics and Computing}, \textbf{17}, 71--80.

\bibitem[{Barber \emph{et~al.}(2012)Barber, Habel, Grasman, Gramacy, Stahel,
  and Sterratt}]{Geometry}
Barber C, Habel K, Grasman R, Gramacy RB, Stahel A, Sterratt DC (2012).
\newblock \emph{\pkg{Geometry}: Mesh Generation and Surface Tesselation}.
\newblock \proglang{R} Package Version 0.3-1,
  \urlprefix\url{http://cran.r-project.org/web/packages/geometry/index.html}.

\bibitem[{Barber \emph{et~al.}(1996)Barber, Dobkin, and
  Huhdanpaa}]{Barber_Etal96}
Barber CB, Dobkin DP, Huhdanpaa H (1996).
\newblock \enquote{The Quickhull Algorithm for Convex Hulls.}
\newblock \emph{ACM Transactions On Mathematicals Software}, \textbf{22},
  469--483.

\bibitem[{Bivand \emph{et~al.}(2012)Bivand, with contributions~by Altman,
  Anselin, Assun\c{c}\~{a}o, Berke, Bernat, Blanchet, Blankmeyer, Carvalho,
  Christensen, Y., Dormann, Dray, Halbersma, Krainski, Legendre, Lewin-Koh, Li,
  Ma, Millo, Mueller, Ono, Peres-Neto, Piras, Reder, Tiefelsdorf, and
  Yu}]{Spdep}
Bivand R, with contributions~by Altman M, Anselin L, Assun\c{c}\~{a}o R, Berke
  O, Bernat A, Blanchet G, Blankmeyer E, Carvalho M, Christensen B, Y C,
  Dormann C, Dray S, Halbersma R, Krainski E, Legendre P, Lewin-Koh N, Li H, Ma
  J, Millo G, Mueller W, Ono H, Peres-Neto P, Piras G, Reder M, Tiefelsdorf M,
  Yu D (2012).
\newblock \emph{\pkg{Spdep}: Spatial Dependence: Weighting Schemes, Statistics
  and Models}.
\newblock \proglang{R} Package Version 0.5-46,
  \urlprefix\url{http://cran.r-project.org/web/packages/spdep/index.html}.

\bibitem[{Bowman and Azzalini(1997)}]{Bowman_Azzalini97}
Bowman AW, Azzalini A (1997).
\newblock \emph{Applied Smoothing Techniques for Data Analysis: the Kernel
  Approach with {S-Plus} Illustrations}.
\newblock Oxford University Press, Oxford.

\bibitem[{Ester \emph{et~al.}(1996)Ester, Kriegel, Sander, and Xu}]{DBSCAN96}
Ester M, Kriegel H, Sander J, Xu X (1996).
\newblock \enquote{A Density Based Algorithm For Discovering Clusters in Large
  Spatial Databases Withe Noise.}
\newblock In \emph{Proceedings of the 2Nd International Conference On Knowledge
  Discovery and Data Mining (KDD-96)}. Aaai Press, Portland.

\bibitem[{Forina \emph{et~al.}(1986)Forina, Armanino, Castino, and
  Ubigli}]{Wine}
Forina M, Armanino C, Castino M, Ubigli M (1986).
\newblock \enquote{Multivariate Data Analysis As a Discriminating Method of the
  Origin of Wines.}
\newblock \emph{Vitis}, \textbf{25}, 189--201.

\bibitem[{Gower(1966)}]{Gower66}
Gower JC (1966).
\newblock \enquote{Some Distance Properties of Latent Root and Vector Methods
  Used in Multivariate Analysis.}
\newblock \emph{Biometrika}, \textbf{53}, 325--328.

\bibitem[{Hartigan(1975)}]{H75}
Hartigan JA (1975).
\newblock \emph{Clustering Algorithms}.
\newblock John Wiley \& Sons, New York.

\bibitem[{Hennig(2010)}]{Fpc}
Hennig C (2010).
\newblock \emph{\pkg{Fpc}: Flexible Procedures for Clustering}.
\newblock \proglang{R} Package Version 2.0-3,
  \urlprefix\url{http://cran.r-project.org/web/packages/fpc/index.html}.

\bibitem[{Hubert and Arabie(1985)}]{Hubert_Arabie85}
Hubert L, Arabie P (1985).
\newblock \enquote{Comparing {P}artitions.}
\newblock \emph{Journal of Classification}, \textbf{2}, 193--218.

\bibitem[{Kaufman and Rousseeuw(1990)}]{Kauf:Rous:1990}
Kaufman L, Rousseeuw PJ (1990).
\newblock \emph{Finding Groups in Data: An Introduction to Cluster Analysis}.
\newblock John Wiley \& Sons, New York.

\bibitem[{Lin \emph{et~al.}(2007)Lin, Lee, and
  Hsieh}]{LinTI:LeeJC:HsiehWJ:2007}
Lin TI, Lee JC, Hsieh WJ (2007).
\newblock \enquote{Robust Mixture Modeling Using the Skew $t$ Distribution.}
\newblock \emph{Statistics and Computing}, \textbf{17}, 81--92.

\bibitem[{Maechler \emph{et~al.}(2005)Maechler, Rousseeuw, Struyf, and
  Hubert}]{Cluster}
Maechler M, Rousseeuw P, Struyf A, Hubert M (2005).
\newblock \enquote{\pkg{Cluster}: Cluster Analysis Basics and Extensions.}
\newblock \proglang{R} package version 1.14.2,
  \urlprefix\url{http://cran.r-project.org/web/packages/cluster/citation.html}.

\bibitem[{McLachlan and Peel(2000)}]{Mclachlan_Peel00}
McLachlan GJ, Peel D (2000).
\newblock \emph{Finite Mixture Models}.
\newblock John Wiley \& Sons, New York.

\bibitem[{McNicholas and Murphy(2008)}]{Mcnicholas_Murphy08}
McNicholas PD, Murphy TB (2008).
\newblock \enquote{Parsimonious Gaussian Mixture Models.}
\newblock \emph{Statistics and Computing}, \textbf{18}, 285--296.

\bibitem[{Menardi(2011)}]{Menardi:2011}
Menardi G (2011).
\newblock \enquote{Density-Based Silhouette Diagnostics for Clustering
  Methods.}
\newblock \emph{Statistics and Computing}, \textbf{21}, 295--308.

\bibitem[{Menardi and Azzalini(2012)}]{Menardi_Azzalini2012}
Menardi G, Azzalini A (2012).
\newblock \enquote{An Advancement in Clustering Via Nonparametric Density
  Estimation.}
\newblock Submitted.

\bibitem[{{R Development Core Team}(2012)}]{R}
{R Development Core Team} (2012).
\newblock \emph{\proglang{R}: A Language and Environment for Statistical
  Computing}.
\newblock R Foundation for Statistical Computing, Vienna, Austria.
\newblock {ISBN} 3-900051-07-0, \urlprefix\url{http://www.R-project.org/}.

\bibitem[{Ray and Lindsay(2005)}]{Ray_Lindsay2005}
Ray S, Lindsay BG (2005).
\newblock \enquote{The Topography of Multivariate Normal Mixtures.}
\newblock \emph{The Annals of Statistics}, \textbf{33}, 2042--2065.

\bibitem[{Rousseeuw(1987)}]{Rousseeuw_87}
Rousseeuw P (1987).
\newblock \enquote{Silhouettes: A Graphical Aid to the Interpretation and
  Validation of Cluster Analysis.}
\newblock \emph{Journal of Computational Applied Mathematics}, \textbf{20},
  53--65.

\bibitem[{Silverman(1986)}]{Silverman86}
Silverman BW (1986).
\newblock \emph{Density Estimation for Statistics and Data Analysis}.
\newblock Chapman and Hall, London.

\bibitem[{Stuetzle(2003)}]{Stuetzle03}
Stuetzle W (2003).
\newblock \enquote{Estimating the Cluster Tree of a Density by Analyzing the
  Minimal Spanning Tree of a Sample.}
\newblock \emph{Journal of Classification}, \textbf{20}, 25--47.

\bibitem[{Stuetzle and Nugent(2010)}]{Stuetzle_Nugent10}
Stuetzle W, Nugent R (2010).
\newblock \enquote{A Generalized Single Linkage Method for Estimating the
  Cluster Tree of a Density.}
\newblock \emph{Journal of Computational and Graphical Statistics},
  \textbf{19}, 397--418.

\bibitem[{Wishart(1969)}]{Wishart69}
Wishart D (1969).
\newblock \enquote{Mode Analysis: A Generalization of Nearest Neighbor Which
  Reduces Chaining Effects.}
\newblock \emph{Numerical Taxonomy}, pp. 282--308.

\end{thebibliography}

\end{document}